\documentstyle[12pt,epsfig,a4]{article}
\newcommand{\btxam}{\mbox{$2.43$}}
\newcommand{\btxaea}{\mbox{$0.20$}}
\newcommand{\btxaeb}{\mbox{$0.25$}}
\newcommand{\btxbm}{\mbox{$3.36$}}
\newcommand{\btxbea}{\mbox{$0.67$}}
\newcommand{\btxbeb}{\mbox{$0.60$}}
\newcommand{\btdm}{\mbox{$0.88$}}
\newcommand{\btdea}{\mbox{$0.31$}}
\newcommand{\btdeb}{\mbox{$0.28$}}
\newcommand{\btm}{\mbox{$8.3 \times 10^{-4}$}}

\newcommand{\tbomhm}{\mbox{$0.49$}}
\newcommand{\btsma}{\mbox{$6.4 \times 10^{-4}$}}

\newcommand{\lamb}{\mbox{$\Lambda_{\mathrm{b}}$}}
\newcommand{\MC}{Monte Carlo}
\newcommand{\cl}{\mbox{$\; \rm{at} \; 90\% \; \rm{CL}$}}
\newcommand{\tpm}{\mbox{$\pm$}}
\newcommand{\tmp}{\mbox{$\mp$}}
\newcommand{\tpc}{\mbox{$\%$}}
\newcommand{\epsb}{\mbox{$\epsilon_b$}}
\newcommand{\epsc}{\mbox{$\epsilon_c$}}
\newcommand{\tanbomh}{\mbox{$\tan \beta / m_{\rm{H}^{\pm}}$}}
\newcommand{\DS}{\mbox{$\rm{D}^{\ast\pm}$}}
\newcommand{\DSS}{\mbox{$\rm{D}^{\ast\ast}$}}
\newcommand{\dza}{\mbox{$\rm{D}^0 \to \rm{K}^- \pi^+$}}
\newcommand{\dzb}{\mbox{$\rm{D}^0 \to \rm{K}^- \pi^+\pi^-\pi^+$}}
\newcommand{\dzc}{\mbox{$\rm{D}^0 \to \rm{K}^- \pi^+\pi^0$}}
\newcommand{\dzd}{\mbox{$\rm{D}^0 \to \rm{K}^0_{\rm{S}} \pi^+\pi^-$}}
\newcommand{\btx}{\mbox{${\mathrm{b}} \to \tau^- \bar{\nu}_{\tau} \rm{X}$}}
\newcommand{\btd}{\mbox{${\mathrm{b}} \to \tau^- \bar{\nu}_{\tau} \DS \rm{X}$}}
\newcommand{\bt}{\mbox{$\rm{B}^- \to \tau^- \bar{\nu}_{\tau}$}}
\newcommand{\bts}{\mbox{${\mathrm{b}} \to \mathrm{s} \nu \bar{\nu}$}}
\newcommand{\btxs}{\mbox{$\rm{B} \to X_s \nu \bar{\nu}$}}
\newcommand{\bbtx}{\mbox{$\rm{BR}(\btx)$}}
\newcommand{\bbtxSM}{\mbox{$\rm{BR}^{\rm{SM}}(\btx)$}}
\newcommand{\bbtd}{\mbox{$\rm{BR}(\btd)$}}
\newcommand{\bbt}{\mbox{$\rm{BR}(\bt)$}}
\newcommand{\bbtSM}{\mbox{$\rm{BR}^{\rm{SM}}(\bt)$}}
\newcommand{\bbts}{\mbox{$\rm{BR}(\bts)$}}

\newcommand{\dst}{\mbox{$\rm{D}_s^{\pm} \to \tau^{\pm} \nu_{\tau}$}}
\newcommand{\xb}{\mbox{$\langle x_{\mathrm b} \rangle$}}
\newcommand{\xc}{\mbox{$\langle x_{\mathrm c} \rangle$}}
\newcommand{\emu}{\mbox{${\mathrm e}/\mu$}}
\newcommand{\bctl}{\mbox{${\mathrm{b,c}} \to \ell \nu_\ell \rm{X}$}}
\newcommand{\btl}{\mbox{${\mathrm b} \to \ell \nu_{\ell} \rm{X}$}}

\newcommand{\btctl}{\mbox{${\mathrm b} \to {\mathrm c} \to 
\ell \nu_\ell \rm{X}$}}
\newcommand{\fcha}{\mbox{$f_{\rm{cha}}$}}
\newcommand{\fpho}{\mbox{$f_{\rm{pho}}$}}
\newcommand{\fneu}{\mbox{$f_{\rm{neu}}$}}
\newcommand{\echa}{\mbox{$E_{\rm{cha}}$}}
\newcommand{\epho}{\mbox{$E_{\rm{pho}}$}}
\newcommand{\eneu}{\mbox{$E_{\rm{neu}}$}}
\newcommand{\evis}{\mbox{$E_{\rm{vis}}$}}
\newcommand{\emiss}{\mbox{$E_{\rm{miss}}$}}

\newcommand{\fextra}{\mbox{$f_{\rm{b}}$}}
\newcommand{\uu}{\mbox{${\mathrm u}\bar{\mathrm u}$}}
\newcommand{\dd}{\mbox{${\mathrm d}\bar{\mathrm d}$}}
\newcommand{\ssq}{\mbox{${\mathrm s}\bar{\mathrm s}$}}
\newcommand{\cc}{\mbox{${\mathrm c}\bar{\mathrm c}$}}
\newcommand{\bb}{\mbox{${\mathrm b}\bar{\mathrm b}$}}

\newcommand{\zbb}{\mbox{$\rm{Z} \to \bb$}}

\newcommand{\ztt}{\mbox{$\rm{Z} \to \tau^+\tau^-$}}
\newcommand{\gev}{\mbox{$\; \rm{GeV}$}}
\def\Rb{\mbox{$R_{\mathrm b}$}}
\def\Rc{\mbox{$R_{\mathrm c}$}}
\def\epsb{\mbox{$\varepsilon_{\mathrm b}$}}
\def\epsbs{\mbox{$\varepsilon_{\mathrm b}^2$}}
\def\epsc{\mbox{$\varepsilon_{\mathrm c}$}}
\def\epscs{\mbox{$\varepsilon_{\mathrm c}^2$}}
\def\epsx{\mbox{$\varepsilon_{\mathrm x}$}}
\def\epsxs{\mbox{$\varepsilon_{\mathrm x}^2$}}
\def\Mcs{\hbox{MeV}/\mbox{$c^2$}}
\def\epem{\mbox{$\hbox{e}^+\hbox{e}^-$}}
\def\Gc{\hbox{GeV}/\mbox{$c$}}
\def\Gcs{\hbox{GeV}/\mbox{$c^2$}}
\def\tanb{\mbox{$\tan\beta$}}
\def\mHpm{\mbox{$m_{{\hbox{\eightrm H}}^\pm}$}}
\def\epemto{\mbox{$\hbox{e}^+\hbox{e}^- \to$}}
\def\Z{\mbox{$\hbox{Z}$}}
\def\mb{\hbox{$m_{\mathrm b}$}}
\def\mc{\hbox{$m_{\mathrm c}$}}
\begin{document}
\font\eightrm=cmr8
\font\ninerm=cmr9
\date{}
\title{ \null\vspace{1cm}
Measurements of\\ 
\bbtx\ and \bbtd\\
and Upper Limits on\\
\bbt\ and \bbts
\vspace{1cm}}
\author{The ALEPH Collaboration$^*)$}

\maketitle

\begin{picture}(160,1)
\put(0,115){\rm ORGANISATION EUROP\'EENNE POUR LA RECHERCHE NUCL\'EAIRE (CERN)}
\put(30,110){\rm Laboratoire Europ\'een pour la Physique des Particules}
\put(125,94){\parbox[t]{45mm}{\tt CERN-EP/2000-126}}
\put(125,88){\parbox[t]{45mm}{\tt 29-Sept-2000}}
\end{picture}

\vspace{.2cm}
\begin{abstract}
\vspace{.2cm}
Inclusive branching ratios involving ${\mathrm b} \to \tau$ transitions
are measured in approximately four million hadronic Z decays collected 
by the ALEPH detector at LEP. The fully-inclusive branching ratio \bbtx\ 
and the semi-inclusive branching ratio \bbtd\ are measured to be 
\mbox{$(\btxam \tpm \btxaea \tpm \btxaeb)\tpc$} and 
\mbox{$(\btdm \tpm \btdea \tpm \btdeb)\tpc$}, in agreement with the 
standard model predictions. Upper limits on the branching fractions 
\bbt\ and \bbts\ are set to \btm\ and \btsma\ at the 90\%~C.L. 
These results allow a 90\% C.L. lower limit of \mbox{$0.40\,(\Gcs)^{-1}$} 
to be set on the $\tanb/\mHpm$ ratio, in the framework of type-II 
two-Higgs-doublet models.
\end{abstract}

\vfill
\centerline{\it Submitted to European Physical Journal C}
\vskip .5cm
\noindent
--------------------------------------------\hfil\break
{\ninerm $^*)$ See next pages for the list of authors}

\eject

\pagestyle{empty}
\newpage
\small
%
\newlength{\saveparskip}
\newlength{\savetextheight}
\newlength{\savetopmargin}
\newlength{\savetextwidth}
\newlength{\saveoddsidemargin}
\newlength{\savetopsep}
\setlength{\saveparskip}{\parskip}
\setlength{\savetextheight}{\textheight}
\setlength{\savetopmargin}{\topmargin}
\setlength{\savetextwidth}{\textwidth}
\setlength{\saveoddsidemargin}{\oddsidemargin}
\setlength{\savetopsep}{\topsep}
%
%
\setlength{\parskip}{0.0cm}
\setlength{\textheight}{25.0cm}
\setlength{\topmargin}{1.7cm}
\setlength{\textwidth}{16 cm}
\setlength{\oddsidemargin}{1.6cm}
\setlength{\topsep}{1mm}
\pretolerance=10000
\centerline{\large\bf The ALEPH Collaboration}
\footnotesize
\vspace{0.5cm}
{\raggedbottom
\begin{sloppypar}
\samepage\noindent
R.~Barate,
D.~Decamp,
P.~Ghez,
C.~Goy,
\mbox{J.-P.~Lees},
E.~Merle,
\mbox{M.-N.~Minard},
P.~Perrodo,
B.~Pietrzyk
\nopagebreak
\begin{center}
\parbox{15.5cm}{\sl\samepage
Laboratoire de Physique des Particules (LAPP), IN$^{2}$P$^{3}$-CNRS,
F-74019 Annecy-le-Vieux Cedex, France}
\end{center}\end{sloppypar}
\vspace{2mm}
\begin{sloppypar}
\noindent
S.~Bravo,
M.P.~Casado,
M.~Chmeissani,
J.M.~Crespo,
E.~Fernandez,
\mbox{M.~Fernandez-Bosman},
Ll.~Garrido,$^{15}$
E.~Graug\'{e}s,
M.~Martinez,
G.~Merino,
R.~Miquel,
Ll.M.~Mir,
A.~Pacheco,
A.~Pascual,
H.~Ruiz
\nopagebreak
\begin{center}
\parbox{15.5cm}{\sl\samepage
Institut de F\'{i}sica d'Altes Energies, Universitat Aut\`{o}noma
de Barcelona, E-08193 Bellaterra (Barcelona), Spain$^{7}$}
\end{center}\end{sloppypar}
\vspace{2mm}
\begin{sloppypar}
\noindent
A.~Colaleo,
D.~Creanza,
M.~de~Palma,
G.~Iaselli,
G.~Maggi,
M.~Maggi,$^{1}$
S.~Nuzzo,
A.~Ranieri,
G.~Raso,$^{23}$
F.~Ruggieri,
G.~Selvaggi,
L.~Silvestris,
P.~Tempesta,
A.~Tricomi,$^{3}$
G.~Zito
\nopagebreak
\begin{center}
\parbox{15.5cm}{\sl\samepage
Dipartimento di Fisica, INFN Sezione di Bari, I-70126
Bari, Italy}
\end{center}\end{sloppypar}
\vspace{2mm}
\begin{sloppypar}
\noindent
X.~Huang,
J.~Lin,
Q. Ouyang,
T.~Wang,
Y.~Xie,
R.~Xu,
S.~Xue,
J.~Zhang,
L.~Zhang,
W.~Zhao
\nopagebreak
\begin{center}
\parbox{15.5cm}{\sl\samepage
Institute of High Energy Physics, Academia Sinica, Beijing, The People's
Republic of China$^{8}$}
\end{center}\end{sloppypar}
\vspace{2mm}
\begin{sloppypar}
\noindent
D.~Abbaneo,
G.~Boix,$^{6}$
O.~Buchm\"uller,
M.~Cattaneo,
F.~Cerutti,
G.~Dissertori,
H.~Drevermann,
R.W.~Forty,
M.~Frank,
T.C.~Greening,
J.B.~Hansen,
J.~Harvey,
P.~Janot,
B.~Jost,
I.~Lehraus,
P.~Mato,
A.~Minten,
A.~Moutoussi,
F.~Ranjard,
L.~Rolandi,
D.~Schlatter,
M.~Schmitt,$^{20}$
O.~Schneider,$^{2}$
P.~Spagnolo,
W.~Tejessy,
F.~Teubert,
E.~Tournefier,
A.E.~Wright
\nopagebreak
\begin{center}
\parbox{15.5cm}{\sl\samepage
European Laboratory for Particle Physics (CERN), CH-1211 Geneva 23,
Switzerland}
\end{center}\end{sloppypar}
\vspace{2mm}
\begin{sloppypar}
\noindent
Z.~Ajaltouni,
F.~Badaud,
G.~Chazelle,
O.~Deschamps,
A.~Falvard,
P.~Gay,
C.~Guicheney,
P.~Henrard,
J.~Jousset,
B.~Michel,
S.~Monteil,
\mbox{J-C.~Montret},
D.~Pallin,
P.~Perret,
F.~Podlyski
\nopagebreak
\begin{center}
\parbox{15.5cm}{\sl\samepage
Laboratoire de Physique Corpusculaire, Universit\'e Blaise Pascal,
IN$^{2}$P$^{3}$-CNRS, Clermont-Ferrand, F-63177 Aubi\`{e}re, France}
\end{center}\end{sloppypar}
\vspace{2mm}
\begin{sloppypar}
\noindent
J.D.~Hansen,
J.R.~Hansen,
P.H.~Hansen,
B.S.~Nilsson,
A.~W\"a\"an\"anen
\begin{center}
\parbox{15.5cm}{\sl\samepage
Niels Bohr Institute, DK-2100 Copenhagen, Denmark$^{9}$}
\end{center}\end{sloppypar}
\vspace{2mm}
\begin{sloppypar}
\noindent
G.~Daskalakis,
A.~Kyriakis,
C.~Markou,
E.~Simopoulou,
A.~Vayaki
\nopagebreak
\begin{center}
\parbox{15.5cm}{\sl\samepage
Nuclear Research Center Demokritos (NRCD), GR-15310 Attiki, Greece}
\end{center}\end{sloppypar}
\vspace{2mm}
\begin{sloppypar}
\noindent
A.~Blondel,$^{12}$
G.~Bonneaud,
\mbox{J.-C.~Brient},
A.~Roug\'{e},
M.~Rumpf,
M.~Swynghedauw,
M.~Verderi,
\linebreak
H.~Videau
\nopagebreak
\begin{center}
\parbox{15.5cm}{\sl\samepage
Laboratoire de Physique Nucl\'eaire et des Hautes Energies, Ecole
Polytechnique, IN$^{2}$P$^{3}$-CNRS, \mbox{F-91128} Palaiseau Cedex, France}
\end{center}\end{sloppypar}
\vspace{2mm}
\begin{sloppypar}
\noindent
E.~Focardi,
G.~Parrini,
K.~Zachariadou
\nopagebreak
\begin{center}
\parbox{15.5cm}{\sl\samepage
Dipartimento di Fisica, Universit\`a di Firenze, INFN Sezione di Firenze,
I-50125 Firenze, Italy}
\end{center}\end{sloppypar}
\vspace{2mm}
\begin{sloppypar}
\noindent
A.~Antonelli,
M.~Antonelli,
G.~Bencivenni,
G.~Bologna,$^{4}$
F.~Bossi,
P.~Campana,
G.~Capon,
V.~Chiarella,
P.~Laurelli,
G.~Mannocchi,$^{5}$
F.~Murtas,
G.P.~Murtas,
L.~Passalacqua,
\mbox{M.~Pepe-Altarelli}$^{24}$
\nopagebreak
\begin{center}
\parbox{15.5cm}{\sl\samepage
Laboratori Nazionali dell'INFN (LNF-INFN), I-00044 Frascati, Italy}
\end{center}\end{sloppypar}
\vspace{2mm}
\begin{sloppypar}
\noindent
A.W. Halley,
J.G.~Lynch,
P.~Negus,
V.~O'Shea,
C.~Raine,
\mbox{P.~Teixeira-Dias},
A.S.~Thompson
\nopagebreak
\begin{center}
\parbox{15.5cm}{\sl\samepage
Department of Physics and Astronomy, University of Glasgow, Glasgow G12
8QQ,United Kingdom$^{10}$}
\end{center}\end{sloppypar}
\vspace{2mm}
\begin{sloppypar}
\noindent
R.~Cavanaugh,
S.~Dhamotharan,
C.~Geweniger,$^{1}$
P.~Hanke,
G.~Hansper,
V.~Hepp,
E.E.~Kluge,
A.~Putzer,
J.~Sommer,
K.~Tittel,
S.~Werner,$^{19}$
M.~Wunsch$^{19}$
\nopagebreak
\begin{center}
\parbox{15.5cm}{\sl\samepage
Kirchhoff-Institut f\"r Physik, Universit\"at Heidelberg, D-69120
Heidelberg, Germany$^{16}$}
\end{center}\end{sloppypar}
\newpage
\vspace{2mm}
\begin{sloppypar}
\noindent
R.~Beuselinck,
D.M.~Binnie,
W.~Cameron,
P.J.~Dornan,
M.~Girone,
N.~Marinelli,
J.K.~Sedgbeer,
J.C.~Thompson,$^{14}$
E.~Thomson$^{22}$
\nopagebreak
\begin{center}
\parbox{15.5cm}{\sl\samepage
Department of Physics, Imperial College, London SW7 2BZ,
United Kingdom$^{10}$}
\end{center}\end{sloppypar}
\vspace{2mm}
\begin{sloppypar}
\noindent
V.M.~Ghete,
P.~Girtler,
E.~Kneringer,
D.~Kuhn,
G.~Rudolph
\nopagebreak
\begin{center}
\parbox{15.5cm}{\sl\samepage
Institut f\"ur Experimentalphysik, Universit\"at Innsbruck, A-6020
Innsbruck, Austria$^{18}$}
\end{center}\end{sloppypar}
\vspace{2mm}
\begin{sloppypar}
\noindent
C.K.~Bowdery,
P.G.~Buck,
A.J.~Finch,
F.~Foster,
G.~Hughes,
R.W.L.~Jones,
N.A.~Robertson
\nopagebreak
\begin{center}
\parbox{15.5cm}{\sl\samepage
Department of Physics, University of Lancaster, Lancaster LA1 4YB,
United Kingdom$^{10}$}
\end{center}\end{sloppypar}
\vspace{2mm}
\begin{sloppypar}
\noindent
I.~Giehl,
K.~Jakobs,
K.~Kleinknecht,
G.~Quast,$^{1}$
B.~Renk,
E.~Rohne,
\mbox{H.-G.~Sander},
H.~Wachsmuth,
C.~Zeitnitz
\nopagebreak
\begin{center}
\parbox{15.5cm}{\sl\samepage
Institut f\"ur Physik, Universit\"at Mainz, D-55099 Mainz, Germany$^{16}$}
\end{center}\end{sloppypar}
\vspace{2mm}
\begin{sloppypar}
\noindent
A.~Bonissent,
J.~Carr,
P.~Coyle,
O.~Leroy,
P.~Payre,
D.~Rousseau,
M.~Talby
\nopagebreak
\begin{center}
\parbox{15.5cm}{\sl\samepage
Centre de Physique des Particules, Universit\'e de la M\'editerran\'ee,
IN$^{2}$P$^{3}$-CNRS, F-13288 Marseille, France}
\end{center}\end{sloppypar}
\vspace{2mm}
\begin{sloppypar}
\noindent
M.~Aleppo,
F.~Ragusa
\nopagebreak
\begin{center}
\parbox{15.5cm}{\sl\samepage
Dipartimento di Fisica, Universit\`a di Milano e INFN Sezione di Milano,
I-20133 Milano, Italy}
\end{center}\end{sloppypar}
\vspace{2mm}
\begin{sloppypar}
\noindent
H.~Dietl,
G.~Ganis,
A.~Heister,
K.~H\"uttmann,
G.~L\"utjens,
C.~Mannert,
W.~M\"anner,
\mbox{H.-G.~Moser},
S.~Schael,
R.~Settles,$^{1}$
H.~Stenzel,
W.~Wiedenmann,
G.~Wolf
\nopagebreak
\begin{center}
\parbox{15.5cm}{\sl\samepage
Max-Planck-Institut f\"ur Physik, Werner-Heisenberg-Institut,
D-80805 M\"unchen, Germany\footnotemark[16]}
\end{center}\end{sloppypar}
\vspace{2mm}
\begin{sloppypar}
\noindent
P.~Azzurri,
J.~Boucrot,$^{1}$
O.~Callot,
S.~Chen,
A.~Cordier,
M.~Davier,
L.~Duflot,
\mbox{J.-F.~Grivaz},
Ph.~Heusse,
A.~Jacholkowska,$^{1}$
F.~Le~Diberder,
J.~Lefran\c{c}ois,
\mbox{A.-M.~Lutz},
\mbox{M.-H.~Schune},
\mbox{J.-J.~Veillet},
I.~Videau,
C.~Yuan,
D.~Zerwas
\nopagebreak
\begin{center}
\parbox{15.5cm}{\sl\samepage
Laboratoire de l'Acc\'el\'erateur Lin\'eaire, Universit\'e de Paris-Sud,
IN$^{2}$P$^{3}$-CNRS, F-91898 Orsay Cedex, France}
\end{center}\end{sloppypar}
\vspace{2mm}
\begin{sloppypar}
\noindent
G.~Bagliesi,
T.~Boccali,
G.~Calderini,
V.~Ciulli,
L.~Fo\`{a},
A.~Giassi,
F.~Ligabue,
A.~Messineo,
F.~Palla,$^{1}$
G.~Sanguinetti,
A.~Sciab\`a,
G.~Sguazzoni,
R.~Tenchini,$^{1}$
A.~Venturi,
P.G.~Verdini
\samepage
\begin{center}
\parbox{15.5cm}{\sl\samepage
Dipartimento di Fisica dell'Universit\`a, INFN Sezione di Pisa,
e Scuola Normale Superiore, I-56010 Pisa, Italy}
\end{center}\end{sloppypar}
\vspace{2mm}
\begin{sloppypar}
\noindent
G.A.~Blair,
G.~Cowan,
M.G.~Green,
T.~Medcalf,
J.A.~Strong,
\mbox{J.H.~von~Wimmersperg-Toeller}
\nopagebreak
\begin{center}
\parbox{15.5cm}{\sl\samepage
Department of Physics, Royal Holloway \& Bedford New College,
University of London, Surrey TW20 OEX, United Kingdom$^{10}$}
\end{center}\end{sloppypar}
\vspace{2mm}
\begin{sloppypar}
\noindent
R.W.~Clifft,
T.R.~Edgecock,
P.R.~Norton,
I.R.~Tomalin
\nopagebreak
\begin{center}
\parbox{15.5cm}{\sl\samepage
Particle Physics Dept., Rutherford Appleton Laboratory,
Chilton, Didcot, Oxon OX11 OQX, United Kingdom$^{10}$}
\end{center}\end{sloppypar}
\vspace{2mm}
\begin{sloppypar}
\noindent
\mbox{B.~Bloch-Devaux},$^{1}$
P.~Colas,
S.~Emery,
W.~Kozanecki,
E.~Lan\c{c}on,
\mbox{M.-C.~Lemaire},
E.~Locci,
P.~Perez,
J.~Rander,
\mbox{J.-F.~Renardy},
A.~Roussarie,
\mbox{J.-P.~Schuller},
J.~Schwindling,
A.~Trabelsi,$^{21}$
B.~Vallage
\nopagebreak
\begin{center}
\parbox{15.5cm}{\sl\samepage
CEA, DAPNIA/Service de Physique des Particules,
CE-Saclay, F-91191 Gif-sur-Yvette Cedex, France$^{17}$}
\end{center}\end{sloppypar}
\vspace{2mm}
\begin{sloppypar}
\noindent
S.N.~Black,
J.H.~Dann,
R.P.~Johnson,
H.Y.~Kim,
N.~Konstantinidis,
A.M.~Litke,
M.A. McNeil,
\linebreak
G.~Taylor
\nopagebreak
\begin{center}
\parbox{15.5cm}{\sl\samepage
Institute for Particle Physics, University of California at
Santa Cruz, Santa Cruz, CA 95064, USA$^{13}$}
\end{center}\end{sloppypar}
\vspace{2mm}
\begin{sloppypar}
\noindent
C.N.~Booth,
S.~Cartwright,
F.~Combley,
M.~Lehto,
L.F.~Thompson
\nopagebreak
\begin{center}
\parbox{15.5cm}{\sl\samepage
Department of Physics, University of Sheffield, Sheffield S3 7RH,
United Kingdom$^{10}$}
\end{center}\end{sloppypar}
\newpage
\vspace{2mm}
\begin{sloppypar}
\noindent
K.~Affholderbach,
A.~B\"ohrer,
S.~Brandt,
C.~Grupen,$^{1}$
A.~Misiejuk,
G.~Prange,
U.~Sieler
\nopagebreak
\begin{center}
\parbox{15.5cm}{\sl\samepage
Fachbereich Physik, Universit\"at Siegen, D-57068 Siegen,
 Germany$^{16}$}
\end{center}\end{sloppypar}
\vspace{2mm}
\begin{sloppypar}
\noindent
G.~Giannini,
B.~Gobbo
\nopagebreak
\begin{center}
\parbox{15.5cm}{\sl\samepage
Dipartimento di Fisica, Universit\`a di Trieste e INFN Sezione di Trieste,
I-34127 Trieste, Italy}
\end{center}\end{sloppypar}
\vspace{2mm}
\begin{sloppypar}
\noindent
J.~Rothberg,
S.~Wasserbaech
\nopagebreak
\begin{center}
\parbox{15.5cm}{\sl\samepage
Experimental Elementary Particle Physics, University of Washington, Seattle, 
WA 98195 U.S.A.}
\end{center}\end{sloppypar}
\vspace{2mm}
\begin{sloppypar}
\noindent
S.R.~Armstrong,
K.~Cranmer,
P.~Elmer,
D.P.S.~Ferguson,
Y.~Gao,
S.~Gonz\'{a}lez,
O.J.~Hayes,
H.~Hu,
S.~Jin,
J.~Kile,
P.A.~McNamara III,
J.~Nielsen,
W.~Orejudos,
Y.B.~Pan,
Y.~Saadi,
I.J.~Scott,
J.~Walsh,
Sau~Lan~Wu,
X.~Wu,
G.~Zobernig
\nopagebreak
\begin{center}
\parbox{15.5cm}{\sl\samepage
Department of Physics, University of Wisconsin, Madison, WI 53706,
USA$^{11}$}
\end{center}\end{sloppypar}
}
\footnotetext[1]{Also at CERN, 1211 Geneva 23, Switzerland.}
\footnotetext[2]{Now at Universit\'e de Lausanne, 1015 Lausanne, Switzerland.}
\footnotetext[3]{Also at Dipartimento di Fisica di Catania and INFN Sezione di
 Catania, 95129 Catania, Italy.}
\footnotetext[4]{Also Istituto di Fisica Generale, Universit\`{a} di
Torino, 10125 Torino, Italy.}
\footnotetext[5]{Also Istituto di Cosmo-Geofisica del C.N.R., Torino,
Italy.}
\footnotetext[6]{Supported by the Commission of the European Communities,
contract ERBFMBICT982894.}
\footnotetext[7]{Supported by CICYT, Spain.}
\footnotetext[8]{Supported by the National Science Foundation of China.}
\footnotetext[9]{Supported by the Danish Natural Science Research Council.}
\footnotetext[10]{Supported by the UK Particle Physics and Astronomy Research
Council.}
\footnotetext[11]{Supported by the US Department of Energy, grant
DE-FG0295-ER40896.}
\footnotetext[12]{Now at Departement de Physique Corpusculaire, Universit\'e de
Gen\`eve, 1211 Gen\`eve 4, Switzerland.}
\footnotetext[13]{Supported by the US Department of Energy,
grant DE-FG03-92ER40689.}
\footnotetext[14]{Also at Rutherford Appleton Laboratory, Chilton, Didcot, UK.}
\footnotetext[15]{Permanent address: Universitat de Barcelona, 08208 Barcelona,
Spain.}
\footnotetext[16]{Supported by the Bundesministerium f\"ur Bildung,
Wissenschaft, Forschung und Technologie, Germany.}
\footnotetext[17]{Supported by the Direction des Sciences de la
Mati\`ere, C.E.A.}
\footnotetext[18]{Supported by the Austrian Ministry for Science and Transport.}
\footnotetext[19]{Now at SAP AG, 69185 Walldorf, Germany.}
\footnotetext[20]{Now at Harvard University, Cambridge, MA 02138, U.S.A.}
\footnotetext[21]{Now at D\'epartement de Physique, Facult\'e des Sciences de Tunis, 1060 Le Belv\'ed\`ere, Tunisia.}
\footnotetext[22]{Now at Department of Physics, Ohio State University, Columbus, OH 43210-1106, U.S.A.}
\footnotetext[23]{Also at Dipartimento di Fisica e Tecnologie Relative, Universit\`a di Palermo, Palermo, Italy.}
\footnotetext[24]{Now at CERN, 1211 Geneva 23, Switzerland.}
%
\setlength{\parskip}{\saveparskip}
\setlength{\textheight}{\savetextheight}
\setlength{\topmargin}{\savetopmargin}
\setlength{\textwidth}{\savetextwidth}
\setlength{\oddsidemargin}{\saveoddsidemargin}
\setlength{\topsep}{\savetopsep}
\normalsize
\newpage
\pagestyle{plain}
\setcounter{page}{1}

\pagestyle{plain}
\pagenumbering{arabic}
\setcounter{page}{1}
\pagestyle{plain}

\section{Introduction}
\label{sec:intro}

Third-generation fermions are involved in both the initial and final 
states of ${\mathrm b} \to \tau$ transitions. A measurement of branching 
ratios pertaining to these transitions can be compared to the standard 
model predictions, yielding direct tests of heavy fermion interactions. 
These interactions are especially sensitive to the mechanism underlying 
the electroweak symmetry breaking, {\it i.e.}, to the origin of mass, 
and allow extensions of the standard model to be constrained in this 
respect. In the study presented here, the branching fractions of the 
following processes are considered.

\begin{figure}[htbp]
\begin{center}
\begin{picture}(160,125)
\put(20,5){\epsfxsize120mm\epsfbox{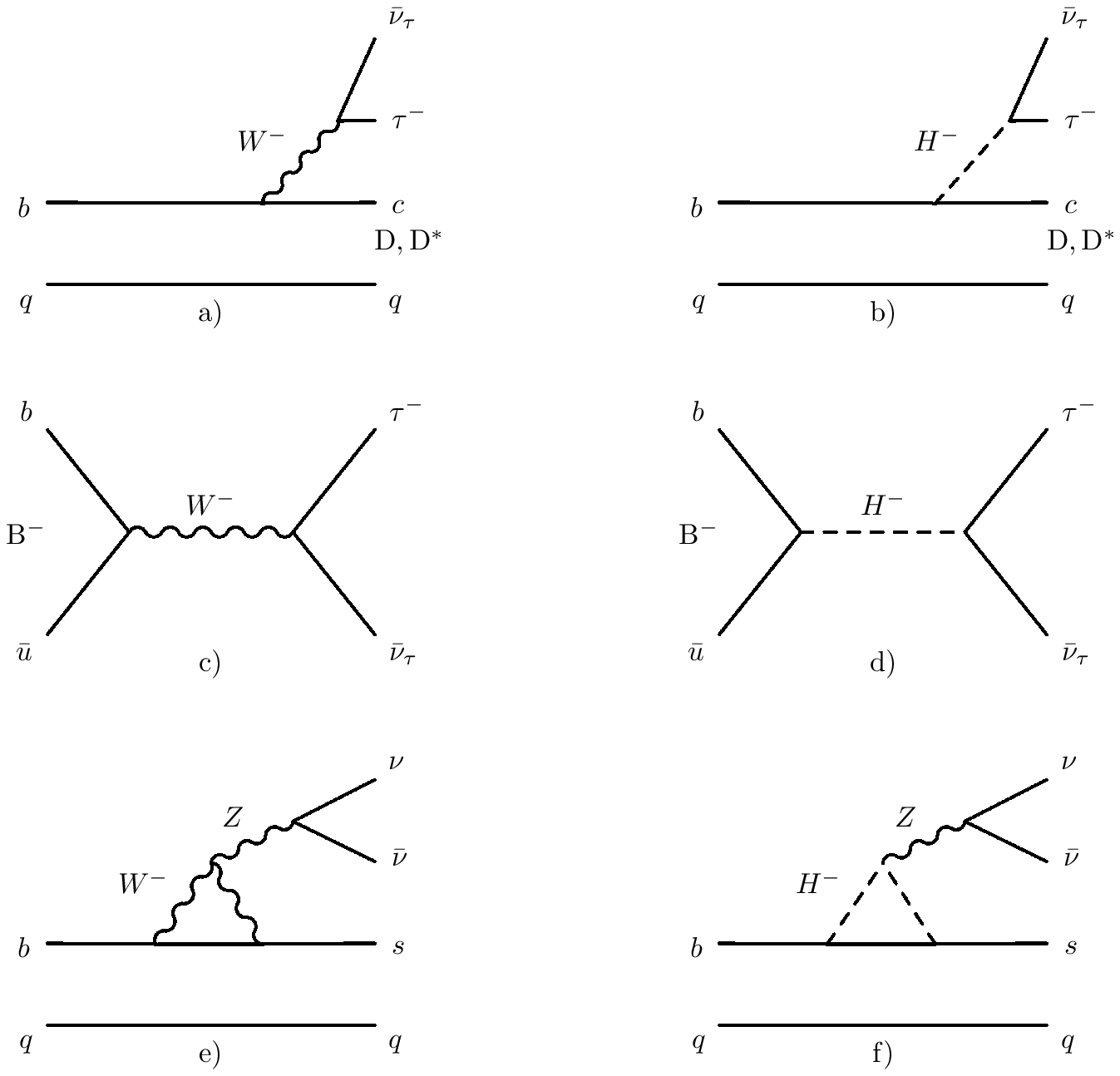}}
\end{picture}
\end{center}
\caption[ ]
{\protect\footnotesize Feynman graphs for the processes \btx, \bt\ and \bts, 
in the standard model (a,c,e) and in two-Higgs-doublet models with the
exchange of a charged Higgs boson (b,d,f).
\label{fig:graphes}} 
\end{figure}

\begin{itemize}
\item
The inclusive branching fraction \bbtx\ can be compared to the 
standard model prediction of 2.30\tpm0.25\tpc~\cite{hqet_btaux1}, 
as computed in the framework of the Heavy Quark Effective Theory (HQET). 
The standard model transition, illustrated in Fig.~\ref{fig:graphes}a, 
could be modified by the exchange of a new charged boson, as shown
in Fig.~\ref{fig:graphes}b with a charged Higgs boson. In any type-II 
two-Higgs-doublet model, such as the minimal supersymmetric extension 
of the standard model, the corresponding contribution to the transition 
amplitude is proportional to 
$\left(\tanb/\mHpm\right)^2$~\cite{higgs_btaux0,higgs_btaux1,higgs_btaux2}, 
where \tanb\ is the ratio of the vacuum expectation values of the two 
Higgs doublets, and \mHpm\ is the mass of the charged Higgs boson.

\item
A similar test can be performed with the semi-inclusive decay ${\rm{b}} \to 
\tau^- \bar{\nu}_{\tau} \mathrm{D}^{\ast+} \mathrm{X}$. In the standard 
model, this branching fraction is expected to be approximately 
1\%~\cite{btauds}. The proportion of ${\mathrm{D}}^\ast$ relative to D 
final states would be reduced in the presence of a charged Higgs 
boson~\cite{higgs_btaux5}. 

\item
The exclusive decay mode \bt\ (Fig.~\ref{fig:graphes}c) has a branching 
fraction predicted to be $7.4 \times 10^{-5}  
\left(f_{\mathrm{B}}/160\,\rm{MeV}\right)^2  
\left(\vert \mathrm{V}_{\mathrm{ub}} \vert /0.004\right)^2$ in the  standard
model~\cite{higgs_btau4}. In type-II two-Higgs-doublet models 
(Fig.~\ref{fig:graphes}d), the rate is enhanced~\cite{higgs_btau4} by the 
multiplicative factor 
$[ m_{{\mathrm{B}}^-}^2 \left(\tanb/\mHpm\right)^2 - 1 ]^2$. 

\item 
Although it does not involve a $\mathrm{b}\to\tau$ transition, the decay 
\bts, with a bran\-ching fraction predicted to be of the order of 
$5\times10^{-5}$ in the standard model (Fig.~\ref{fig:graphes}e), can 
also be exploited to put constraints on a variety of extensions of
the standard model (Fig.~\ref{fig:graphes}f) as advocated in 
Refs.~\cite{bsnunu_theory,bsnunu_theory_b}.
\end{itemize}

The measurements of the branching fractions of these four processes, \btx, 
\btd, \bt\ and \bts, are performed with all data collected by the ALEPH 
detector between 1991 and 1995, at centre-of-mass energies at and around 
the Z resonance, corresponding to approximately four million hadronic Z 
decays. As described in previous ALEPH studies applied to smaller data 
samples~\cite{aleph_btaux1,aleph_btaux2}, a single technique is used 
to extract the four branching ratios. The initial-state b quark is 
identified by means of standard b-tagging algorithms, and the 
final-state $\tau$ lepton (or the two neutrinos in the case of \bts) 
is identified by the missing energy carried away by the neutrinos. 

The energy-flow and b-tagging algorithms are briefly described in 
Section~\ref{sec:aleph-detector} along with properties of the 
ALEPH detector relevant for the present study. The selection 
algorithms aimed at rejecting background events with a large 
missing energy are presented in Section~\ref{sec:analysis-method}. 
The method to estimate the residual background yields is detailed in 
Section~\ref{sec:corr-mc}. The branching ratios \bbtx\ and \bbtd\ are 
determined in Section~\ref{sec:meast-bbtx}, and the limits on \bbt\ and 
\bbts\ are extracted in Section~\ref{sec:limits}. 
Finally, an alternative measurement of \bbtx\ with opposite-sign 
di-leptons in the final state is presented in Section~\ref{sec:meast-dilepton}.
The results are interpreted in the framework of type-II two-Higgs-doublet 
models in Section~\ref{sec:typeII} and summarized in 
Section~\ref{sec:conclusion}.

\parskip= 0.25cm plus .05cm
\section{The ALEPH Detector}
\label{sec:aleph-detector}

A detailed description of the ALEPH detector can be found in 
Ref.~\cite{aleph_detector1}, and of its performance in 
Ref.~\cite{aleph_detector2}. Charged particles are detected in
the central part, consisting of a precision silicon vertex detector,
a cylindrical drift chamber and a large time projection chamber, measuring
altogether  up to 31 space points along the charged particle trajectories.
 A 1.5 T axial magnetic field is provided by a superconducting solenoidal 
coil. Charged particle transverse momenta are reconstructed with a 
1/$p_{T}$ resolution of $\left( 6\times 10^{-4}  \bigoplus 5 \times 
10^{-3}/p_{T}\right)~(\Gc)^{-1}$.
In the following, {\it good} tracks are defined as charged particle
tracks reconstructed with at least four hits in the time projection
chamber, originating from within a cylinder of length 20\,cm
and radius 2\,cm coaxial with the beam and centred at the nominal
collision point, and with a polar angle with respect to the beam
such that $\vert \cos\theta \vert < 0.95$. 

Jets originating from b quarks are identified with a lifetime b-tagging
algorithm~\cite{aleph_btag}, which takes advantage of the three-dimensional 
impact parameter resolution of charged particle tracks. For tracks with 
two space points in the silicon vertex detector ({\it i.e.}, $\vert 
\cos\theta \vert < 0.7$), this resolution can be parametrized as 
$(25+95/p)~\mu$m, with $p$~in~\Gc.

In addition to its r\^ole as a tracking device, the time projection
chamber also measures the specific energy loss by ionization 
${\mathrm d}E/{\mathrm d}x$. It allows low momentum electrons to be 
separated from other charged particle species by more than three 
standard deviations.

Electrons (and photons) are also identified by the characteristic longitudinal
and transverse developments of the associated showers in the electromagnetic 
calorimeter, a 22 radiation length thick sandwich of lead planes and 
proportional wire chambers with fine read-out segmentation. The
relative energy resolution achieved is $0.18/\sqrt{E}$ ($E$ in GeV)
for isolated electrons and photons.

Photon conversions to \epem\ in the detector material are identified 
as a pair of oppositely charged identified electrons satisfying the 
following conditions: {\it (i)} the distance of closest approach to the
beam of the two reconstructed tracks is larger than 2~mm; {\it (ii)} the 
distance between the two tracks at their point of closest approach 
is smaller than 2\,cm in space; {\it (iii)} the position of this point is
consistent with a material boundary; and {\it (iv)} the invariant mass 
is smaller than 20~\Mcs, when calculated as for an $\epem$ pair coming 
from this point of closest approach. 

Muons are identified by their characteristic penetration pattern in the 
hadron calorimeter, a 1.5 m thick yoke interleaved with 23 layers of 
streamer tubes, together with two surrounding double-layers of muon chambers. 
In association with the electromagnetic calorimeter, the hadron calorimeter 
also provides a measurement of the hadronic energy with a relative resolution 
of $0.85/\sqrt{E}$ ($E$ in GeV).

Taus are identified by the missing energy carried away by their
decay neutrinos. The total visible energy is measured with an energy-flow
reconstruction algorithm which combines all the above measurements,
supplemented by the energy detected at low polar angle (down to 24 mrad 
from the beam axis) by two additional electro\-magnetic calorimeters, used 
for the luminosity determination. The relative resolution on the total 
visible energy varies between $0.60/\sqrt{E}$ for high multiplicity 
final states and $0.25/\sqrt{E}$ for final states of low multiplicity 
without neutral hadrons. In addition to the total visible-energy 
measurement, the energy-flow reconstruction algorithm also provides 
a list of reconstructed objects, classified as charged particles, photons 
and neutral hadrons, and called {\it energy-flow particles} in the 
following.

\parskip= 0.15cm plus .05cm
\section{Event selections}
\label{sec:analysis-method}

A ${\mathrm b} \to \tau$ transition followed by the decay of the $\tau$ 
always produces two energetic $\nu_{\tau}$'s. Such transitions can 
therefore be identified in $\epemto\ \bb$ events on the basis 
of a large measured missing energy. The argument holds as well for 
\bts\ decays, which can be selected along the same lines as 
${\mathrm b} \to \tau$ transitions. In this section, the algorithms
aimed at rejecting background events leading to large measured missing 
energy are described. An alternative selection, based on the presence of
two identified leptons with opposite electric charge, is described in 
Section~\ref{sec:meast-dilepton}.

\subsection{Preselection}
\label{sec:presel}

Approximately four million hadronic Z decays are selected in the data 
collected between 1991 and 1995 at energies at and around the Z 
resonance when the standard criteria (at least five good tracks 
carrying at least 10\% of the centre-of-mass energy~\cite{aleph_ztonu})
are applied. To keep only two-jet events well contained in the detector 
acceptance, the polar angle 
is required to satisfy  \mbox{$|\cos \theta_{\rm{thrust}} |< 0.7$} to match
the acceptance of the vertex detector, and the thrust value must exceed 
0.85. 

Each event is then divided in two hemispheres with respect to the 
plane perpendi\-cular to the thrust axis. In each hemisphere, the missing 
energy $E_{\rm{miss}}^{1,2}$ is defined as the difference between 
the expected true energy $E_{\rm{true}}^{1,2}$ and the measured visible 
energy $E_{\rm{vis}}^{1,2}$. The latter is determined from the total energy
of all energy-flow particles contained in that hemisphere, while the 
former is estimated from the centre-of-mass energy $\sqrt{s}$ and 
with energy-momentum conservation: 
$E_{\rm{true}}^{1,2} = ({s + m_{1,2}^2 - m_{2,1}^2}) /2\sqrt{s}$, 
where $m_1$ and $m_2$ are the measured invariant masses in 
the two hemispheres~\cite{napoleon_duarte}.

The main background to the final states searched for in this analysis
consists of $\Z \to \bb$ or $\cc$ events followed by a semi-leptonic
b or c decay into an electron or a muon (hereafter called lepton and 
denoted $\ell$), with a large missing energy carried away by the 
neutrino $\nu_\ell$. This background can be considerably reduced
by rejecting hemispheres in which a lepton is identified. The standard 
lepton identification~\cite{aleph_hvfl_nim} is not used here, but is 
replaced by much looser criteria. A good track is tagged as an electron 
either if the ${\mathrm d}E/{\mathrm d}x$ is compatible with that of an 
electron and incompatible with that of a pion, or if the transverse and 
longitudinal profiles of the associated electromagnetic shower are compatible 
with those of an electron. Electrons and positrons originating from identified 
photon conversions are not considered in this process. Similarly, a good 
track is tagged as a muon if it is associated to a few hits either in the 
last layers of the hadron calorimeter or in the two layers of muon chambers. 
A lepton identification efficiency over 95\% is achieved for electrons 
(muons) with a momentum in excess of 1~(2)\,\Gc.

In addition, to reduce the contamination from Z decays into lighter quark
pairs (\uu, \dd, \ssq\ and \cc) in which a large missing energy is 
faked due to finite detector resolution effects, the final sample 
is enriched in $\Z \to \bb$ events with b tagging. The corresponding
criteria are dependent on the signal final state and are described 
in the following subsections.

\parskip= 0.20cm plus .05cm
\subsection{The 
{\mbox{\boldmath ${\mathrm{b}} \to \tau^- \bar{\nu}_{\tau} \rm{X}$}} 
final state}
\label{sec:btx}

For the \btx\ final state selection, a hemisphere is kept only 
if the opposite hemisphere is tagged as arising from a b quark. 
Each good track in the opposite hemisphere is assigned a probability 
of originating from the primary interaction point, on the basis of 
its impact parameter significance. The confidence level $\alpha^{\rm{hemi}}$ 
that all good tracks in that hemisphere come from the primary interaction 
point, determined under the assumption that the individual probabilities 
are uncorrelated~\cite{aleph_btag}, is required to be smaller than 1\%. 

Residual backgrounds like \ztt\ decays, two-photon processes or 
beam-gas interactions are likely to yield a large missing energy in the 
final state and might therefore bias the analysis. These background sources 
are reduced down to negligible levels by requiring at least seven good 
tracks and a total missing energy smaller than 50\,GeV, with almost no 
additional loss of the signal.

  
\subsection{The 
{\mbox{\boldmath $\rm{B}^- \to \tau^- \bar{\nu}_{\tau}$}}
and 
{\mbox{\boldmath ${\mathrm{b}} \to \mathrm{s} \nu \bar{\nu}$}}
final states}
\label{sec:btnbts}

The same b-tagging technique as for the \btx\ selection 
is applied but the residual backgrounds are rejected 
in a slightly different manner. The opposite hemisphere is 
required to contain at least six good tracks, and its missing 
energy must be smaller than 25\,GeV. This latter cut reduces the 
effect on the missing-energy distribution in the signal hemisphere, 
which is expected to extend towards larger values than in the previous 
case. 

\subsection{The 
{\mbox{\boldmath ${\mathrm{b}} \to \tau^- \bar{\nu}_{\tau} \DS \rm{X}$}}
final state}
\label{sec:btd}

The \btd\ final state selection is primarily based on an 
exclusive reconstruction of a \DS, very similar to that 
performed in Ref.~\cite{aleph_charmed_meson} although with 
slightly tighter cuts as a result of the larger statistics 
accumulated until 1995. 

Candidate $\DS$'s are searched for in the decay channel 
$\DS \to \rm{D}^0 \pi_{\rm{soft}}$, followed by \dzb, \dza, \dzc\ 
or \dzd, selected as follows. 

Neutral pions are identified as photon pairs with an invariant mass 
compatible with the $\pi^0$ mass and a $\chi^2$ probability in excess 
of 1\%. Candidate ${\mathrm K}_{\mathrm S}^0$'s are identified as pairs 
of oppositely charged particle tracks (assumed to be charged pions) 
forming a secondary vertex at least 2\,cm away from the primary interaction 
point, and with a reconstructed invariant mass  within 5\,\Mcs\ of the 
${\mathrm K}_{\mathrm S}^0$ mass.  Candidate ${\mathrm K}^\pm$'s are 
identified as good tracks of momentum in excess of 3\,\Gc\ and with a 
${\mathrm d}E/{\mathrm d}x$ compatible with that of a charged kaon within 
two standard deviations. In addition the angle $\theta^\ast$ between the 
kaon momentum direction, evaluated in the $\rm{D}^0$ rest frame, and the 
$\rm{D}^0$ line-of-flight must satisfy $|\cos \theta^\ast| < \rm{0.9}$. 

The reconstructed invariant mass of the candidate $\rm{D}^0$'s is 
required to be within 15\,\Mcs\ (50\,\Mcs) of the $\rm{D}^0$ mass  
for the first (last) two decay channels. The ratio of the $\rm{D}^0$ 
energy to the beam energy is required to be between 0.25 and 0.50. 
In addition, the reconstructed $\rm{D}^0$ vertex must be incompatible 
with the primary interaction point by at least two standard deviations. 

Each reconstructed $\rm{D}^0$ is associated with a low momentum pion 
($p_{\pi_{\rm{soft}}}<4.2\,\Gc$) to form a \DS. The reconstructed 
invariant mass difference $m_{\rm{D}^{\ast\pm}}-m_{\rm{D}^0}$ is required 
to be within 20\,\Mcs\ of the nominal mass difference. 

Finally, the \DS\ hemispheres are tagged as arising from a b quark 
by requiring the presence of an additional good track in a cone 
of half-angle $30^\circ$ around the $\rm{D}^0$ momentum direction.
This track must be incompatible with originating from the primary 
interaction point by more than three standard deviations, and its 
electric charge must be opposite to that of the $\pi_{\rm{soft}}$.
No b tagging needs therefore be applied to the opposite hemisphere.

\subsection{Event yields}
\label{sec:mc-sample}

The expected event yields were determined with the following samples 
of events, simulated with {\tt JETSET}~\cite{jetset72,jetset73}, and
subsequently processed through the full {\tt GEANT}~\cite{geant} simulation 
of the detector:
\begin{itemize}
\item approximately four million hadronic Z decays for the background
estimate;
\item over 40\,000 \bb\ events, each containing a \btx, with the 
$\tau$ polarization deter\-mined in the limit of the free-quark 
model~\cite{hqet_btaux1} for $\mb = 4.8\,\Gcs$ and $\mc = 1.4\,\Gcs$;
\item about 20\,000 \bb\ events with at least one \bt\ and 100\% pola\-rized
$\tau$'s ;
\item almost 20\,000 \bb\ events with at least one \bts, generated with 
the \btxs\ decay probability given in Ref.~\cite{bsnunu_theory}, in which 
the new-physics parameters were conservatively chosen to minimize the expected 
selection efficiency. The model was improved with a realistic 
$\mathrm{X}_{\mathrm{s}}$ invariant mass distribution, described as a 
Gaussian of mean and variance 1.35 and 0.60\,\Gcs~\cite{bsnunu_theory_b}, 
supplemented by two peaks at the K and $\mathrm{K}^\ast$ 
masses~\cite{mxs-distribution} with branching fractions of $0.06\pm0.02$ and 
$0.29\pm0.01$.
\end{itemize}
In all these  samples, the polarization of the $\tau$'s was properly 
taken into account in the simulation of their decay kinematics~\cite{koralz}, 
and events containing a $\Lambda_{\mathrm b}$ were reweighted according 
to the measured $\Lambda_{\mathrm b}$ polarization~\cite{aleph_lambpol},
so that the correct distribution of the $\nu_\tau$ momenta, and therefore
the missing energy, be obtained. 

The numbers of hemispheres expected to be selected from the various background
and signal processes and the numbers of hemispheres selected in the data are 
listed in Table~\ref{tab:event-yield}.

\begin{table}[htbp]
\begin{center}
\caption{\footnotesize Numbers of hemispheres selected in the data in the 
four different final states. Also indicated are the numbers of hemispheres 
expected from the various background and signal processes. For the latter, 
the standard model branching ratio values were assumed. The standard model 
branching ratios and the selection efficiencies are indicated in 
brackets.
\label{tab:event-yield}}
\vspace{3mm}
\begin{tabular}{|l|r|r|r|} \hline
Final state selection & $\tau^-\bar\nu_\tau\mathrm{X}$ 
                      & $\tau^-\bar\nu_\tau\mathrm{D}^{\ast\pm}\mathrm{X}$ 
                      & $\tau^-\bar\nu_\tau$, $\mathrm{s}\nu\bar\nu$ \\ \hline 
\multicolumn{4}{|c|}{Data} \\ \hline
Hemispheres selected       
                    & 166342       & 1464          & 156910 \\ \hline\hline
\multicolumn{4}{|c|}{Simulation} \\ \hline
Hemispheres expected       
                    & 162456       & 1538          & 153093 \\ \hline
\multicolumn{4}{|c|}{Background (details)} \\ \hline
u, d, s             & 9215         & 1172          & 8855   \\ \hline
b, c without leptonic decay 
                    & 138118       & 80            & 129886  \\ \hline
b, c with leptonic decay 
                    & 10645        & 180           & 10091  \\ \hline
\multicolumn{4}{|c|}{Signal (details)} \\ \hline
\btx\ (2.3\%)       & (12\%) 4478  & --            & 4247   \\ \hline
\btd\ (1.0\%)       & --           & (0.34\%) 81   & --     \\ \hline
${\mathrm{b}} \to \tau^- \bar{\nu}_{\tau} \rm{X \; without}\ \DS$
                    & --           & 25            & --     \\ \hline
\bt\ $(7.4\times 10^{-5})$ 
                    & --           & --            & (8.1\%) 4   \\ \hline 
\bts\ $(5.0\times 10^{-5})$ 
                    & --           & --            & (8.8\%) 10   \\ \hline 
\end{tabular}
\end{center}
\end{table}

\parskip= 0.15cm plus .05cm
The measurement of the various branching fractions relies on the observation 
of an excess of events at large \emiss\ in the missing-energy distribution 
of the data with respect to that of the simulated background. Given 
the limited number of events expected from the various signal processes
(Table~\ref{tab:event-yield}), it is essential to have a detailed
understanding of all uncertainties affecting the background distribution 
both in shape and normalization (Section~\ref{sec:corr-mc}) and of the 
signal selection efficiencies (Section~\ref{sec:meast-bbtx}).

\section{Background estimate}
\label{sec:corr-mc}

The missing-energy distribution of the background to $ {\mathrm b} \to 
\tau$ and $ {\mathrm b} \to {\mathrm s} \nu \bar\nu$ transitions
depends on detector performance in three areas, {\it (i)} the visible-energy 
reconstruction for hemispheres with {\it a priori} no missing energy 
such as Z hadronic decays into light-quark pairs; {\it (ii)} the effectiveness 
of the lepton veto to reject background from ${\mathrm b, c} \to \ell \nu_\ell 
{\mathrm X}$; and {\it (iii)} the performance of the b-tagging. To minimize 
the influence of possible inaccuracies in the detector simulation, these 
three quantities are derived from the data and subsequently used to re-weight 
the simulated events.

\subsection{The visible-energy calibration}
\label{sec:corr-mc-emiss}

As mentioned above, the accuracy of the visible-energy reconstruction 
can be studied with a sample of events in which no significant missing 
energy is expected. Such a sample is selected in a way identical 
to that followed in Sections~\ref{sec:presel} and~\ref{sec:btx} except 
that the b-tagging criterion is replaced by a b-rejection one requiring 
$\alpha^{\rm{hemi}}>0.5$. It yields fractions of 88.7\%, 10.5\% and 0.8\% 
for (\uu,\dd,\ssq), \cc\ and \bb\ events, respectively.

The visible energy \evis\ is the sum of the energy \echa\ of the charged 
particles reconstructed in the central tracker, the energy \epho\ of the 
photons detected in the electromagnetic calorimeter, and the energy \eneu\ 
of the neutral hadrons, the reconstruction of which makes use of all 
identification and energy measurement capa\-bi\-lities of the detector
(Section~\ref{sec:aleph-detector}). These three contributions to the 
visible energy may therefore be affected by different systematic 
uncertainties and must be calibrated separately. The hemispheres were 
thus binned in a grid according to the fractional contributions to the 
visible energy of the charged, photon and neutral hadron components 
(\echa/\evis, \epho/\evis, \eneu/\evis). The accuracy of the detector 
simulation was examined by comparing the observed and simulated missing 
energy in each of the bins.

A noticeable inaccuracy of the simulation was found in hemispheres with a 
large proportion of neutral hadronic energy. This disagreement with the 
data could lead to a large systematic uncertainty in the final 
result. Moreover, the neutral hadronic energy is expected to be 
smaller in \bb\ events than in other hadronic Z decays. It was therefore 
decided to reject all hemispheres for which \eneu\ is in excess of 
7\,GeV, both in this calibration procedure and in the selections
described in the previous section, preserving 69.6\tpc\ of the 
events in the data and 68.7\tpc\ in the simulation.

The residual differences between the data and the simulation were 
corrected by scaling \echa, \epho\ and \eneu\ by \fcha, \fpho\ 
and \fneu, respectively, in the simulated events. The values of these 
calibration factors were obtained by minimizing the following $\chi^2$,
\begin{equation}
\chi^2 = \sum_{i}\left\{ 
\langle E_{\rm{vis},i}^{\rm{data}} \rangle 
- \left[
\fcha\langle E_{{\rm{cha}},i}^{\rm{MC}}\rangle
+\fpho\langle E_{{\rm{pho}},i}^{\rm{MC}}\rangle
+\fneu\langle E_{{\rm{neu}},i}^{\rm{MC}}\rangle\right] \right\}^2/\sigma^2_i,
\end{equation}
where the mean energy values are computed in each bin $i$, 
$E_{{\rm{vis}},i}^{\rm{data}}$ is the total visible energy measured 
in that bin, the $E_{{\rm{xxx}},i}^{\rm{MC}}$ values are the charged, 
photonic and neutral hadronic energies expected in that bin, and $\sigma_i$ 
is the uncertainty due to the limited statistics of the event samples. 
The result of the fit is shown in Table~\ref{tab:emiss-factors}. While 
\fcha\ and \fpho\ are compatible with unity, a sizeable calibration is 
found to be still necessary for the neutral hadronic energy, despite  
the cut at 7\,GeV. This effect, already reported in Ref.~\cite{aleph_btaux1}, 
results from an inadequacy of the simulation of nuclear interactions. 
\begin{table}[htbp]
\begin{center}
\caption{\footnotesize Calibration parameters (see text for a detailed 
description) for \evis.
\label{tab:emiss-factors}}
\vspace{3mm}
\begin{tabular}{|l||c|} \hline\hline
\fcha    &   1.002 \tpm\ 0.001           \\ \hline
\fpho    &   1.000 \tpm\ 0.002           \\ \hline
\fneu    &   0.936 \tpm\ 0.010           \\ \hline\hline
\fextra  &   0.999 \tpm\ 0.001           \\ \hline\hline
\end{tabular}
\end{center}
\end{table}

As another consequence, the \eneu\ resolution is better in the simulation 
than in the data. In addition to the rescaling, the simulated neutral 
hadronic energy was therefore smeared on an event-by-event basis, leading to a 
$(1.1 \pm 1.6)\%$ relative deterioration of the \eneu\ resolution. This
smearing was performed in order to equalize the observed and expected 
numbers of events with a reconstructed \emiss\ in excess of 14\,GeV.
Indeed, in events where no real missing energy is expected, such a 
large reconstructed \emiss\ value is often due to a loss of neutral 
hadronic energy.

\begin{figure}[t]
\begin{picture}(160,75)
\put(0,-2){\epsfxsize160mm\epsfbox{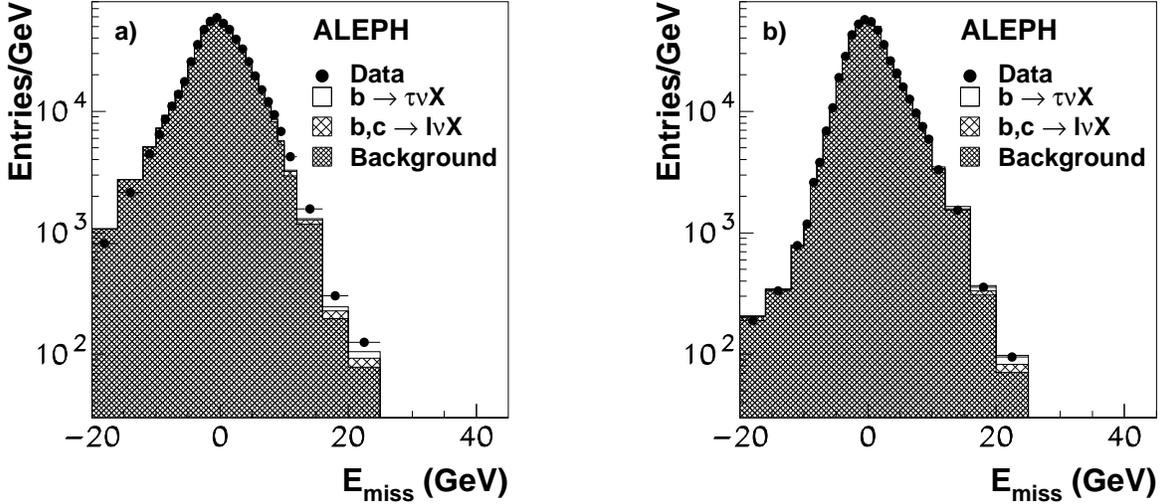}}
\end{picture}
\caption[ ]
{\protect\footnotesize Reconstructed missing-energy distributions 
using the light-quark tag and \emu\ veto (a) before calibration; 
(b) after missing-energy recalibration and applying the cut
$\eneu<7\,$GeV.
\label{plot-emisslight}} 
\end{figure}

\parskip= 0.15cm plus .05cm
The distributions of the reconstructed missing energy in the 
control sample are displayed in Fig.~\ref{plot-emisslight}, both 
for data and simulated events, before and after the calibration 
procedure. The agreement is found to be adequate over the whole 
spectrum, and in particular at large missing energy where the 
background to the signal of interest in this paper has to be evaluated.

Finally, due to the specific hadronization and fragmentation of b-quark 
jets, the sharing between the charged, photonic and neutral hadronic energy 
may lead to an overall visible-energy shift different from that observed for 
lighter quark species. The visible energy of hemispheres tagged as arising 
from a b quark was therefore rescaled by an overall factor \fextra, 
determined from the b-tagged sample of Section~\ref{sec:btx} so as to 
match the peak positions (which are not affected by events with truly 
missing energy) of the expected and observed visible-energy distributions. 
This b-specific calibration factor is found to be compatible with unity
(Table~\ref{tab:emiss-factors}).

\subsection{The lepton rejection effectiveness}

In order to estimate the effectiveness of the lepton veto, aimed at rejecting
$\Z \to \bb$ or $\cc$ events followed by a semi-leptonic b or c decay 
into an electron or a muon, the veto was applied to unbiased, pure,
lepton data and simulated samples. 

A 97.6\% pure electron sample was obtained by selecting hadronic events with a 
photon conversion, as described in Section~\ref{sec:aleph-detector}, in which 
only one of the two particles has to be identified as an electron. Photon 
conversions are well suited to study the electron veto since the hadronic 
environment is very similar to that of semi-leptonic heavy-flavour decays. 
The electron-veto effectiveness is thus given by the probability, after 
background subtraction, that the second particle be identified as an electron 
by the lepton-veto criteria. However, this probability depends on {\it (i)} 
the electron momentum, the spectrum of which is different in photon 
conversions and in semi-leptonic b/c decays; and {\it (ii)} the number of 
wires used for the ${\mathrm d}E/{\mathrm d}x$ measurement, which tends to 
be smaller in photon conversions than in semi-leptonic b/c decays due to 
shared hits with the first electron of the converted pair. Therefore, the 
electron identification probability was mapped as a function of the electron 
momentum and the number of hits associated to the corresponding track, 
both for the data and the simulated events with photon conversions. The 
ratio of these two maps was then used to re-weight the simulated b/c events 
with semi-leptonic decays.

\begin{figure}[b]
\begin{picture}(160,75)
\put(0,-2){\epsfxsize160mm\epsfbox{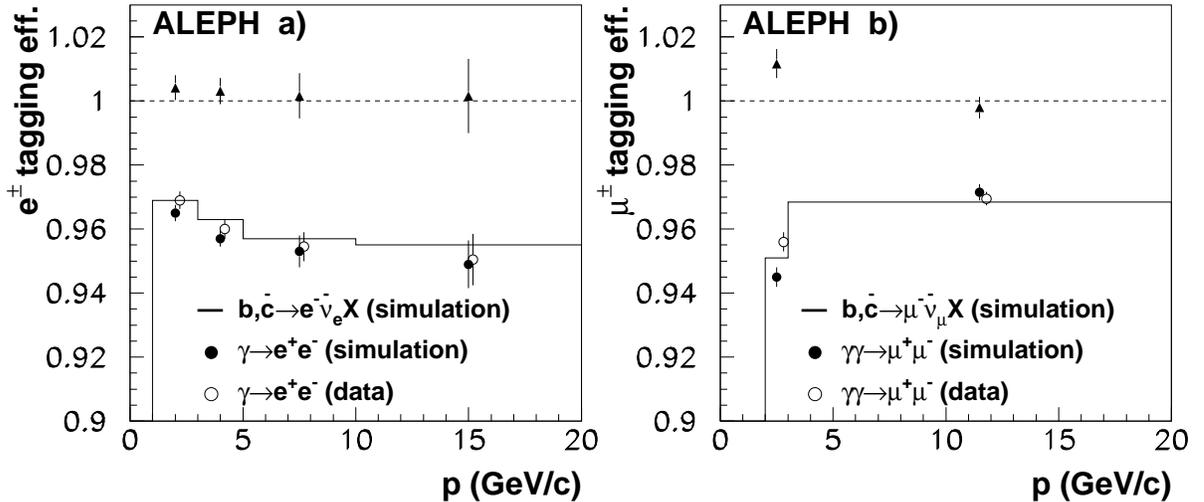}}
\end{picture}
\caption[ ]
{\protect\footnotesize Lepton identification efficiencies (a) for electrons 
and (b) for muons as a function of momentum in the control samples (see text)
for the data (open circles) and the simulation (full circles), and their ratio
(triangles). The histograms show the identification efficiency for 
simulated b/c events with semi-leptonic decays.
\label{fig:leptoneff}}
\end{figure}

For the muons, a sample originating from the $\gamma\gamma \to \mu^+\mu^-$ 
process was selected. The absence of hadronic environment is less important 
because the identification is achieved with the external layers of the hadron 
calorimeter and the muon chambers. A 97.1\% pure sample of low-momentum 
muons was obtained by requiring exactly two good tracks (accompanied by no 
neutral energy-flow particle) with opposite electric charge, one of which 
being identified as a muon, a total momentum less than 30\,\Gc\ 
and an invariant mass smaller than 2\,\Gcs. Above 3\,\Gc, muons are 
expected to traverse entirely the hadron calorimeter and the muon chambers, 
making the identification efficiency independent of momentum. The 
$\gamma\gamma \to \mu^+\mu^-$ sample was therefore supplemented by Z decays 
into muon pairs to determine the latter. As above, and after background 
subtraction, the muon-veto effectiveness was given by the probability 
that the second particle be identified as a muon by the lepton-veto 
criteria as a function of momentum, both in the data and in the 
simulation. The ratio was then used to re-weight the simulated b/c 
events with semi-leptonic decays.

The identification probability for electrons and muons, determined from 
these samples, is displayed as a function of the lepton momentum in 
Figs.~\ref{fig:leptoneff}a and~\ref{fig:leptoneff}b, for data and simulation, 
and is compared to the identification efficiency for simulated b/c 
events with semi-leptonic decays. The data-to-simulation ratio, applied as 
a correction factor to the latter, is consistently found to be close to 
unity.

\parskip .30cm plus .10cm
\subsection{The b-tagging efficiency}
\label{sec:bb-tagging}

The efficiency of the b-tagging criterion can be determined directly 
from the data using the double tag method~\cite{aleph_btag}, so as to 
minimize the systematic uncertainties related to the limited knowledge 
of b-hadron production and decay. The small correlation between the  
b-tagging probabilities in the two hemispheres is negligible for the 
present purpose. 

From the knowledge of the fractions 
\Rb\ and \Rc\ of hadronic Z decays into \bb\ and \cc, the efficiencies 
\epsb\ and \epsc\ of the b-tagging criterion on b and c hemispheres can be 
measured in the data by comparing the fraction $f_1$ of hemispheres 
which pass a given $\alpha^{\rm{hemi}}$ cut with the fraction $f_2$ of events 
in which both hemispheres pass the same cut. These fractions are given by
\begin{eqnarray}
f_1 & = & \Rb \epsb + \Rc \epsc + (1-\Rc-\Rb) \epsx, \cr
f_2 & = & \Rb \epsbs + \Rc \epscs + (1-\Rc-\Rb) \epsxs,
\label{eq:btag}
\end{eqnarray}
where \epsx\ is the hemisphere tagging efficiency for light-quark 
hemispheres, the value of which is determined directly with the data as
explained in Ref.~\cite{aleph_btag}. The difference between the observed and 
expected \epsx\ values for a given $\alpha^{\rm{hemi}}$ cut arises from 
from different impact parameter resolutions in the data and in the simulation.
This difference was estimated from the fraction of hemispheres satisfying 
$\alpha_{\rm{neg}}^{\rm{hemi}} < 0.01$, where $\alpha_{\rm{neg}}^{\rm{hemi}}$ 
is the hemisphere probability recomputed with only those tracks having negative
impact parameter (and reversing the impact parameter sign). Such tracks nearly 
always originate from the primary vertex, and account for resolution  
differences. The relatively small contribution of long-lived particles (such 
as $\rm{K}^0_S$ or $\Lambda^0$) is taken from the simulation.


With the cut $\alpha^{\rm{hemi}} < 0.01$, \epsx\ was found to be $2.46\%$,
{\it i.e.}, slightly larger in the data than predicted by the simulation 
by a factor of $1.17 \pm 0.11$. With the \epsx\ value obtained this way, 
Eq.~(\ref{eq:btag}) was solved for \epsb\ and \epsc. The  tagging efficiencies 
for b (56.7\%) and c (14.8\%) hemispheres were also found to be larger in 
the data by factors of $1.03 \pm 0.01$ and $1.03 \pm 0.03$, respectively, 
and were used to re-weight the simulated events accordingly. 

\section{Measurement of 
\mbox{\boldmath  $\rm{b} \to \tau^- \bar{\nu}_{\tau} \rm{X~and~b} \to 
\tau^- \bar{\nu}_{\tau} \DS \rm{X}$}}
\label{sec:meast-bbtx}

\begin{figure}[htbp]
\begin{picture}(160,75)
\put(0,-2){\epsfxsize165mm\epsfbox{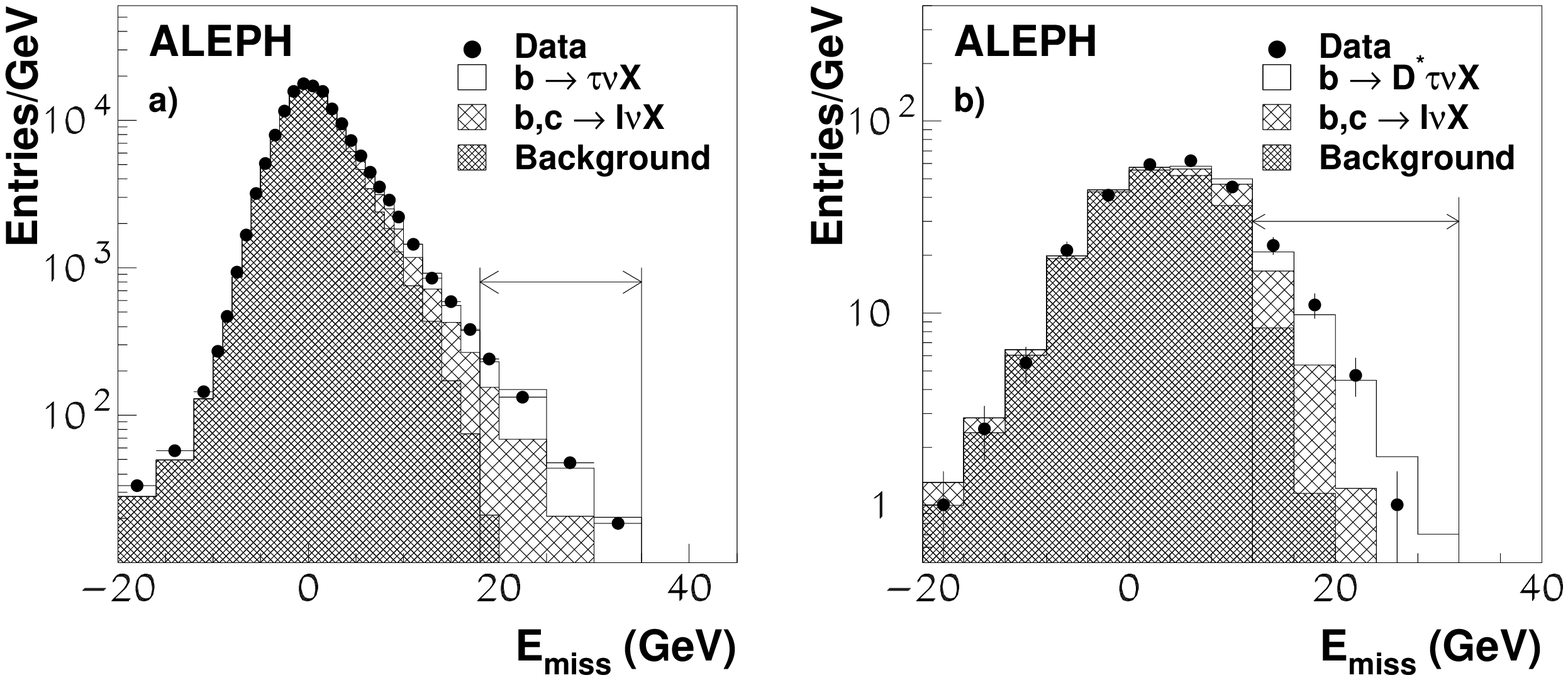}}
\end{picture}
\caption[ ]
{\protect\footnotesize Missing-energy distributions in the \btx\ (a) and
the \btd\ (b) final states, for the data (dots) and the simulation
(histograms). The latter is subdivided in {\it (i)} the fitted signal 
contribution
(empty histogram); {\it (ii)} the contribution from b and c semi-leptonic
decays (light hatching); and {\it (iii)} the residual background
(dark hatching). Also indicated are the \emiss\ intervals considered 
for the branching ratio measurements (Section~\ref{sec:extraction}).
\label{fig:plotbtaunux}}
\end{figure}

The $\rm{b} \to \tau^- \bar{\nu}_{\tau} \rm{X~and~b} \to 
\tau^- \bar{\nu}_{\tau} \DS \rm{X}$ branching fractions can be 
determined by evaluating the excess of events over the background 
expected at large missing energy from the two selections 
(Sections~\ref{sec:btx} and~\ref{sec:btd}). The \emiss\ distributions 
in the \btx\ and \btd\ final states are displayed in 
Figs.~\ref{fig:plotbtaunux}a and~\ref{fig:plotbtaunux}b, for the data 
and for the re-weighted simulation (Section~\ref{sec:corr-mc}), 
normalized to the number of observed events.

\subsection{Branching ratio extraction}
\label{sec:extraction}                  

An excess of events at large \emiss\ is indeed observed over the background 
in both distributions. The corresponding branching ratios were extracted 
with a binned-likelihood fit of the expected to the observed missing-energy 
distributions, keeping the normalization to the number of events observed.
The fit was performed in \emiss\ intervals chosen so as to {\it (i)} 
minimize their total uncertainty, defined as the quadratic sum of the 
statistical and the systematic contributions (Section~\ref{sec:btx-syst}); 
and {\it (ii)} make the measurements statistically independent of that 
of the \bt\ branching fraction. The optimal intervals, indicated in 
Fig.~\ref{fig:plotbtaunux}, are found to be 18-35\,GeV and 
12-35\,GeV, respectively.

The numbers of events in these intervals, observed in the data and 
expected from both the background and the signal processes, are 
displayed in Table~\ref{tab:event-meast-bbtx}. In particular, 
the cascade decay $\rm{b} \to \rm{D}^-_s \rm{X}$ with 
$\rm{D}^-_s\to\tau^-\bar{\nu}_{\tau}$, which yields an \emiss\ spectrum 
similar to that of the decay \btx, is included in the backgrounds to this 
signature.

These numbers yield measured branching fractions of
\vskip -.1cm
\begin{eqnarray*}
\bbtx\phantom{\DS} & = & \left[ \btxam \pm \btxaea \rm{(stat.)} \pm \btxaeb {\rm (syst.)}
\right]\%, \\ 
 & \rm{and} & \\ 
\bbtd & = & \left[ \btdm \pm \btdea \rm{(stat.)} \pm \btdeb {\rm (syst.)}
\right]\%,
\end{eqnarray*}

\vskip .3cm
\noindent
in agreement with the standard model predictions of $(2.30 \pm 0.25)\%$ and 
approximately 1\%, respectively. The systematic uncertainties, also indicated 
above, are discussed in the next section.

\begin{table}[ht]
\begin{center}
\caption{\footnotesize Numbers of events observed in the data at 
large \emiss\ in the \btx\ and \btd\ final states. Also indicated
are the events expected from the various background and 
signal processes. For the latter, the fitted branching ratio 
value was assumed. The corresponding selection efficiencies are indicated 
in brackets.
\label{tab:event-meast-bbtx}}
\vspace{3mm}
\begin{tabular}{|l|r|r|} \hline 
Final state selection 
& \multicolumn{1}{|c|}{$\tau^-\bar\nu_\tau\mathrm{X}$} 
& \multicolumn{1}{|c|}{$\tau^-\bar\nu_\tau\mathrm{D}^{\ast\pm}\mathrm{X}$}
\\ \hline
\emiss\ interval & \multicolumn{1}{|c|}{[18,35]\,GeV}   
                  & \multicolumn{1}{|c|}{[12,35]\,GeV}    \\ \hline 
\multicolumn{3}{|c|}{Data}                                \\ \hline
Hemispheres selected  & 2094          & 162               \\ \hline 
\multicolumn{3}{|c|}{Expected background (details)}       \\ \hline
u, d, s                             & 17            &  2  \\ \hline
b, c with leptonic decay            & 1001          &  51 \\ \hline
$\rm{b} \to \rm{D}^-_s \rm{X}$, $\rm{D}^-_s \to \tau^-\bar\nu_\tau$      
                                    & 214             & --  \\ \hline
b, c without $\ell$, $\rm{D}^-_s$   & $84$        & 36  \\ \hline
\multicolumn{3}{|c|}{Expected signal (details)}           \\ \hline
\btx                & (2\%) 778           & --                  \\ \hline
\btd                & --            & (0.26\%) 63                  \\ \hline
${\mathrm{b}} \to \tau^- \bar{\nu}_{\tau} \rm{X \; without} \DS$
                    & --            & 10                  \\ \hline
\end{tabular}
\end{center}
\end{table}

\subsection{Systematic uncertainties}
\label{sec:btx-syst}

The measured branching ratios \bbtx\ and \bbtd\ can be affected by several 
sources of systematic effects on the signal selection efficiencies and
by residual inaccuracies of the background simulation. The relevant sources
common to \btx\ and \btd\ are addressed in turn below, and their effects 
on the branching fractions are summarized in Table~\ref{tab:btaux_syst}.

\begin{figure}[ht]
\begin{picture}(160,110)
\put(25,-1){\epsfxsize110mm\epsfbox{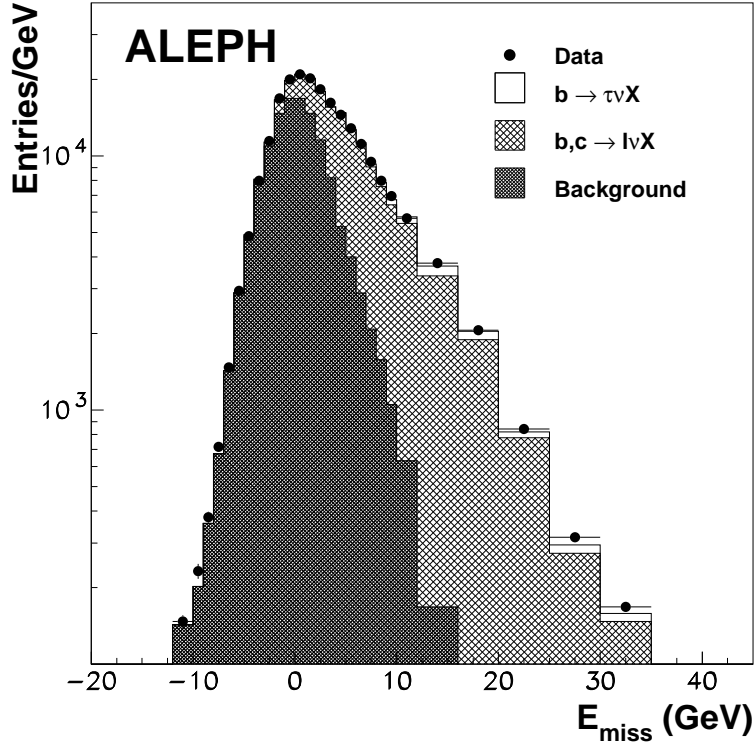}}
\end{picture}
\caption[ ]
{\protect\footnotesize \emiss\ distributions obtained for the data (dots) 
and the simulation (histogram) from the selection described in 
Section~\ref{sec:btx} with the lepton-veto criteria reversed. 
\label{fig:plotbtaunux_l}}
\end{figure}
                                                
\begin{enumerate}

\item The missing-energy distributions of \btx\ and \bctl\ depend on 
the expe\-rimentally determined quantities \xb, \xc, BR(\btl) and BR(\btctl).
However, the small, residual differences between the $\emiss$ distributions 
in the data and the simulation, obtained by reversing the lepton-veto 
criteria (Fig.~\ref{fig:plotbtaunux_l}), were found to be well accounted 
for by the measured uncertainties on their values. The whole extraction 
procedure of the branching fractions was therefore repeated by varying 
these quantities within their uncertainties, so as to evaluate the 
corresponding systematic effects. Similarly, the branching ratio 
dependence on the \DSS\ content in b decays was accounted for according 
to Ref.~\cite{Vub}.

\item The b-fragmentation modelling affects the b-energy spectrum, 
and therefore also the missing-energy spectrum in  \btx\ and \bctl. 
The effect of its knowledge was estimated in a similar manner by using 
the four different fragmentation functions described in 
Ref.~\cite{lep_ew_hv_1994}, adjusted to reproduce the measured 
value of \xb. The corresponding systematic uncertainty was defined 
as the largest change in the measured branching fractions.

\item The $\bar\nu_\tau$ energy spectrum in the b rest frame also depends 
on whether it is computed with HQET or with the spectator 
model~\cite{hqet_btaux1}. To reflect this difference, the neutrino 
spectrum was distorted with a weight depending linearly on the neutrino 
energy such that the averaged energy changes by $\pm10\%$, and the 
resulting change in the measured branching fractions noted.

\item The $\nu_\tau$ energy spectrum depends on the $\tau$ polarization 
and, for events containing a \lamb, on the $\lamb$ polarization. These
polarizations were varied according to the measured uncertainty on the latter,
and to the difference between the HQET and the spectator model prediction 
for the former, to determine their effects on the branching fractions.

\item The uncertainty on $\rm{b} \to \rm{D}^-_s \rm{X}$, followed
by \dst~\cite{nancy} was propagated to the branching 
fractions. In this process, the missing-energy spectrum is expected to be 
similar to that of the signal. 

\item The small, residual differences between the calibrated missing-energy 
distributions for light-quark events, the b-tagging efficiencies 
and the lepton-veto effectiveness in data and simulation 
(Section~\ref{sec:corr-mc}) were all attributed to systematic effects 
and translated as such to the branching fraction determination.

\item The visible-energy calibration procedure 
(Section~\ref{sec:corr-mc-emiss}) is tuned with light-quark events, and
may not be entirely accurate for \bb\ events in the signal region. 
In this region, the missing energy is found in the simulation 
to mainly originate from mis-reconstructed neutral hadrons. 
It is observed that \bb\ events yield more hemispheres with a large 
neutral hadronic energy in the data than in the simulation. This excess is 
expected to increase the residual background in the signal region by 
$\sim 20\%$, treated as an additional systematic uncertainty.
\end{enumerate}
\noindent
\parskip .15cm plus .05cm
Two additional systematic effects, specific to the \btd\ final state were 
also identified. First, the combinatorial background was estimated in 
that case from the sidebands around the ${\rm D}^0$ peak, and used to 
derive the corresponding uncertainty on \bbtd. Second, the b-tagging 
criterion specific to this channel (Section~\ref{sec:btd}) yields a 
different efficiency in data and simulation, which affects \bbtd.

\begin{table}[htbp]
\caption{\footnotesize Systematic uncertainties (in $\%$) for \bbtx\ and \bbtd.
\label{tab:btaux_syst}}
\vspace{2mm}
\begin{center}
\begin{tabular}{|l|c|c|c|} \hline
\multicolumn{1}{|c|}{Source}                 & \bbtx      & \bbtd     \\ 
\hline
\xb=0.702 \tpm\ 0.008~\cite{lep_ew_hv_1998}   &  \tmp0.12  &  \tmp0.15 \\ 
\xc=0.487 \tpm\ 0.008~\cite{lep_ew_hv_1998}   &  \tmp0.01  &  \tmp0.01 \\   
BR(\btl)=10.56 \tpm\ 0.21 \tpc~\cite{simon}   &  \tmp0.05  &  \tmp0.10 \\ 
BR(\btctl)=7.98 \tpm\ 0.22\tpc~\cite{simon}   &  \tmp0.01  &  \tmp0.03 \\ 
\hline
BR($\rm{b} \to \DSS,\DS\pi$)
=25 \tpm\ 7\tpc~\cite{Vub}                      &  \tpm0.03  &  \tpm0.06 \\
\hline
b-fragmentation 
modelling~\cite{lep_ew_hv_1994}              &  \tmp0.11  &  \tmp0.12 \\
\hline
\btx\ decay modelling                        &  \tpm0.06  &  \tpm0.06 \\       
\hline
BR(\dst)=5.79 \tpm\ 1.94\tpc~\cite{nancy}    &  \tmp0.08  &  \tmp0.07 \\     
\hline
$\langle\rm{P}_{\tau}\rangle=
-0.735\pm0.03$~\cite{hqet_btaux1}            &  \tpm0.02  &  \tpm0.01 \\     
$\langle\rm{P}(\lamb)\rangle
=-0.31^{+0.22}_{-0.19}
\pm0.08$~\cite{aleph_lambpol}                &  \tmp0.06  &  \tmp0.01 \\   
\hline
b-tagging efficiency                         &  \tpm0.06  &  \tpm0.02 \\
$\mu$-identification efficiency              &  \tpm0.06  &  \tpm0.07 \\      
e-identification efficiency                  &  \tpm0.08  & \tpm0.08  \\    
Visible-energy calibration                   &  \tmp0.05  &  \tmp0.04 \\ 
\hline
\rule{0pt}{4.0mm}
\emiss\ in \bb\ events                       &  \tpm0.06  &  \tpm0.05 \\   
\hline       
Combinatorial background               &  --  &  \tpm0.03 \\ 
\hline       
Total systematic uncertainty                 & \tpm\btxaeb & \tpm\btdeb \\ 
\hline 
\end{tabular}
\end{center}
\end{table}

\section{Upper limits on \mbox{\boldmath{\bt}} and \mbox{\boldmath{\bts}}}
\label{sec:limits}

The \bt\ and \bts\ branching ratios are predicted to be too small 
in the standard model to be measured with only four million hadronic
Z decays. However, upper limits can be set on these branching fractions
to constrain possible extensions of the standard model, such as type-II
two-Higgs-doublet models. Since \bbt\ and \bbts\ can be affected differently
depending on the new physics considered, the upper limits are conservatively 
estimated here for each process separately.

\begin{figure}[htbp]
\begin{picture}(160,100)
\put(25,-1){\epsfxsize110mm\epsfbox{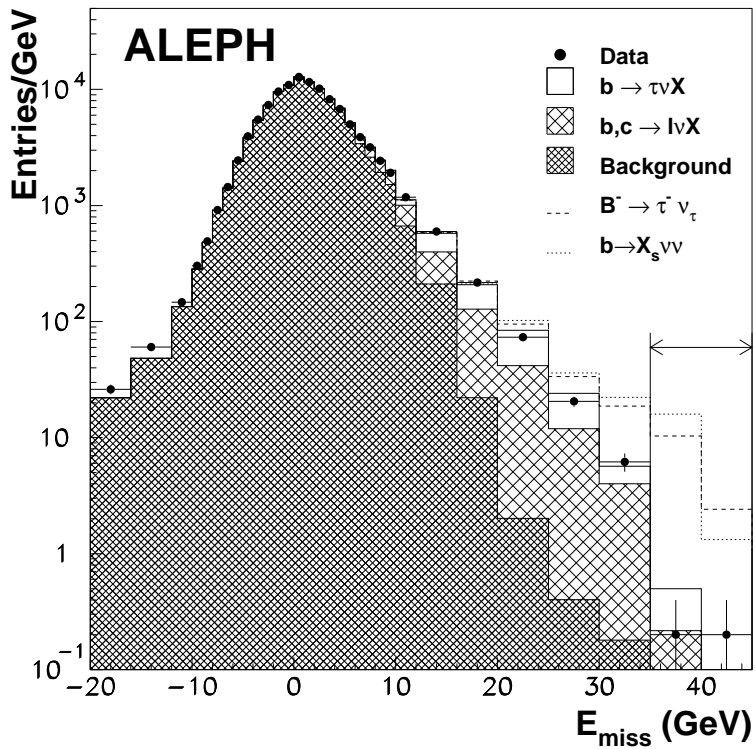}}
\end{picture}
\caption[ ]
{\protect\footnotesize Missing-energy distributions in the \bt\ and \bts\ 
final state selection, for the data (dots) and for the simulated
background (full histogram). Also indicated are the contributions
of the \bt\ (dashed histogram) and \bts\ (dotted histogram) processes
with a branching ratio of 1\%. The arrow shows the region in which the 
two limits are calculated.
\label{fig:plotsearch}}
\end{figure}

The missing-energy distribution of the events selected as described in 
Section~\ref{sec:btnbts} is displayed in Fig.~\ref{fig:plotsearch}, and 
is compared to that of the background. Also indicated in 
Fig.~\ref{fig:plotsearch} is the expected enhancement at large missing 
energy, should either the \bt\ or the \bts\ branching ratio be 1\%.

\noindent
No background subtraction was performed, making the validity of the
results presented here unaffected by possible systematic uncertainties
related to the knowledge of the resi\-dual background, largely suppressed
by a lower cut on the missing energy at 35\,GeV. The latter cut was optimized 
so as to maximize the expected 90\% C.L. upper limit, evaluated with simple 
event counting and in the absence of new physics~\cite{aleph_standard_higgs}, 
on the \bt\ and \bts\ branching ratios. The numbers of events, observed 
in the data and expected from background and signal, are displayed in 
Table~\ref{tab:nbbtnu} in three missing-energy intervals. 
Two events with a missing energy in excess of 35\,GeV were observed, with 
$2.5\pm 1.6$ events expected from all background processes. In absence of 
any systematic uncertainty, the 90\% C.L. upper limits on the \bt\ and \bts\  
branching fractions are found to  be $8.1 \times 10^{-4}$ and 
$6.2 \times 10^{-4}$, respectively.

\begin{table}[htbp]
\begin{center}
\caption{\footnotesize Numbers of events observed in the data in 
three different \emiss\ intervals. Also indicated are the events 
expected from the various background and signal processes. 
For the latter, a branching ratio of 1\% was assumed.
\label{tab:nbbtnu}}
\vspace{3mm}
\begin{tabular}{|l|l|l|l|} \hline
\emiss\ interval & \multicolumn{1}{|c|}{[30, 35]\,GeV}
                 & \multicolumn{1}{|c|}{[35, 40]\,GeV}
                 & \multicolumn{1}{|c|}{$> 40$\,GeV} \\ \hline
\multicolumn{4}{|c|}{Data} \\ \hline
Selected hemispheres & 31            & 1             &   1           \\  \hline
\multicolumn{4}{|c|}{Simulation} \\ \hline
Expected hemispheres & 37.0$\pm 2.7$ & $2.5\pm 1.6$  &   $< 1$       \\  \hline
\multicolumn{4}{|c|}{Background (detail)} \\ \hline
\btx                 &   7.2         &   1.4         &   --          \\
b, c with leptonic decay 
                     &  28.9         &   1.1         &   --          \\
Other Backgrounds    &   0.9         &   --          &   --          \\ \hline
\multicolumn{4}{|c|}{Signal (BR=1\%)}\\ \hline
\bt                  & 76.9          & 53.1          & 12.7 \\
\bts                 & 91.0          & 78.6          & 6.6  \\ \hline
\end{tabular}
\end{center}
\end{table}

However, these limits are affected by the uncertainty on the expected 
fraction of \bt\ and \bts\ events with such a large missing energy or, 
almost equivalently, with such a large value of $x_{\mathrm b}$. 
(Most of these events 
are characterized by a value of $x_{\mathrm b}$ in excess of 0.9.) This 
fraction was determined~\cite{aleph_xb} to be $0.146^{+0.025}_{-0.021}$, 
which translates to an uncertainty of 15\% on the number of events expected. 
In addition, the fraction of $\rm{B}^-$ in \zbb\ events is 
known~\cite{aleph_bminusfraction} to be $(38.9\pm1.3)\%$. The latter uncertainty 
of 3\% on the number of events expected affects only the limit on the \bt\ 
branching ratio. These two uncertainties were taken into account following 
the method of Ref.~\cite{cousins}, yielding the limits 
\begin{eqnarray*}
\bbt  & < & \btm,    \\
\bbts & < & \btsma, 
\end{eqnarray*}
at the $90\%$ confidence level.
\section{Measurement of \mbox{\boldmath \bbtx} with di-leptons}
\label{sec:meast-dilepton}   

\indent An alternative method of measuring \bbtx\ was developed with 
events where both the $\tau$ and the accompanying D decay to e or $\mu$.
Hence, the signature used to tag the signal events is a pair of leptons
(e, $\mu$) of opposite sign in a jet.
The background, originating from $\rm{b} \to \rm{c} \ell^- \bar{\nu}_{\ell}$ 
followed by $\rm{c} \to \rm{q} \ell^+ \nu_{\ell}$, 
is about 20 times larger than the signal.
Signal and background are therefore separated on the basis of their 
different kinematic properties. 
Although this method is statistically less powerful than that based
on missing energy, 
it represents an interesting cross-check since it is based on a 
complementary sample of events and sources of systematic uncertainties 
are largely different. 
Indeed, the main contribution to the systematic error comes from 
the uncertainty on the product 
\mbox{BR($\rm{b} \to \ell \bar{\nu}_{\ell} \rm{c}$, ${\rm c} \to \ell \nu_\ell 
\rm{q}$)}
and on the double charm decay rates, 
$\rm{B} \to \rm{D_s D (X)}$ and $\rm{B} \to \rm{D^0 D (X)}$.

The presence of three neutrinos in the decay chain of signal events is 
the main difference between signal and background.
As a consequence, $\btx$ decays give larger missing energy,
softer lepton spectrum and smaller charged multiplicity
to the jet containing the lepton candidates.
These different kinematic properties of the various 
categories of events are used to separate the signal from the background. 
A multivariate analysis technique with a multilayered neural network (NN) is 
used to obtain the best discriminating power.   

\subsection{Event selection}

\indent Electron and muon identification follows the standard 
criteria~\cite{aleph_hvfl_nim}, with two refinements for muons. Firstly, 
the muon momentum cut is lowered from $3\gev/c$ to $2.2\gev/c$ 
to increase the acceptance for the signal ($2.2\gev/c$ is the minimum 
momentum for a muon to reach the muon chambers) and secondly, any track 
``shadowed'' by another track is rejected in the di-lepton selection.
Two tracks are said to be shadowing if they share in  common either hits
in the last ten layers of the hadron calorimeter or at least one 
three-dimensional muon chamber hit. This latter cut allows a good 
description of the background to be obtained in the sample of events with 
two muons in the same jet, as shown in Ref.~\cite{tesis}. To  obtain 
the di-lepton sample used for the training of the neural network, two 
additional cuts on the invariant mass $M_{\ell\ell}$ of the lepton pair 
are applied.
\begin{itemize}
\item Electron pair invariant masses are required to be greater 
than $0.3\gev/c^2$ in order to reject $\gamma$ conversions or Dalitz 
decays of $\eta$, $\pi^0$.
\item Electron pairs and muon pairs are required to have an invariant mass
smaller than $2.5\gev/c^2$ to avoid pairs coming from \mbox{$\rm{J}/\psi$} 
decays. 
\end{itemize}
After these selection cuts, the simulated di-lepton sample 
consists of 451 signal events, 7834 $\rm{b} \to \ell$ background events 
and 5249 di-leptons coming from hadrons mis-identified as leptons and 
light-hadron decays.  

\subsection{Analysis method and background estimation}

Jets are reconstructed by the JADE clustering algorithm~\cite{jet} 
with a $y_{cut}$ of 0.0044~\cite{aleph_hvfl_nim}. The following
variables are selected as input to the NN: the missing energy in the
hemisphere of the jet containing the lepton pair,  the momenta of 
the leptons boosted to the reconstructed  b-hadron rest frame~\cite{tesis}, 
the invariant mass of the lepton pair, the transverse 
momenta of the leptons with respect to the jet direction, the 
total energy and the charged energy of the jet containing the lepton pair, 
the  fraction of the jet charged energy carried by the leptons 
and the number of charged particles in the jet.   
             
The main background consists of di-leptons coming from 
$\rm{b} \to \rm{c} \ell, \; \rm{c} \to \ell' \nu_{\ell'}$ decays.
To determine the  product 
BR($\rm{b} \to \ell \bar{\nu}_{\ell} \rm{c}) \times \rm{BR}(c \to \ell' 
\nu_{\ell'})$,
the semi-leptonic branching fraction 
BR($\rm{b} \to \ell \bar{\nu}_{\ell} \rm{X}$) 
is fixed to the LEP average value~\cite{simon},
while the  $\rm{b} \to \rm{c} \to \ell'$ fraction is determined
from the same sample of di-lepton events as follows:
a neural network with the input variables described above is 
trained to separate $\rm{b} \to \rm{c} \ell \to \ell \ell' \rm{X}$ 
decays from all other processes; and the $\rm{b} \to \rm{c} \to \ell$ 
fraction is extracted by fitting the simulated neural network output 
distribution to that of the data. No attempt at an evaluation of the 
systematic uncertainties on this fraction was made.

Another important background consists of leptons from light-ha\-dron decays 
or of hadrons mis-identified as leptons.  
A control sample of same sign di-leptons is used to test the
accuracy of the simulation. This sample has a composition 
similar to that of the background of the analysis, since one 
lepton candidate is always fake or coming from the decay
of a light-flavoured hadron.

The measured values~\cite{doublec} for double charmed b decays 
($\rm{B} \to \rm{D_s D (X)}$, $\rm{B} \to \rm{D^0 D (X)}$)
are included in the background estimate as well. 

\subsection{Branching ratio estimation with a neural network}

\begin{figure}[htbp]
\begin{picture}(160,90)
\put(-5,-2){\epsfxsize170mm\epsfbox{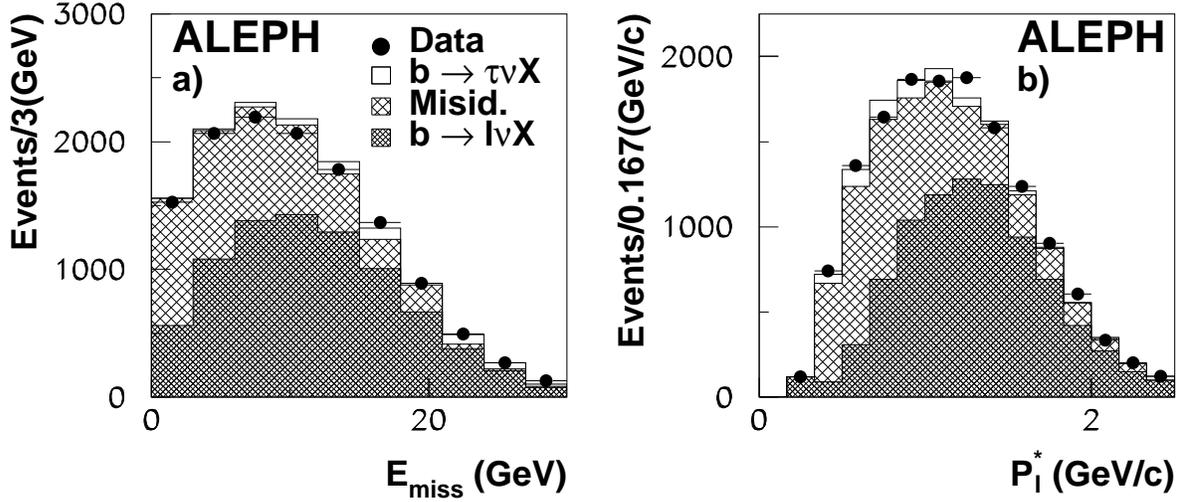}}
\end{picture}
\caption[ ]
{\protect\footnotesize Data and normalized \MC\ distributions
for (a) the missing energy of the leptonic hemisphere
and (b) the momentum of one lepton boosted to the reconstructed 
b-hadron rest frame. The normalization of the simulated distribution 
for signal events corresponds to the \bbtx\ value obtained from the fit.
\label{fig:comp}}
\end{figure}

\begin{figure}[htbp]
\begin{picture}(160,85)
\put(-5,-2){\epsfxsize180mm\epsfbox{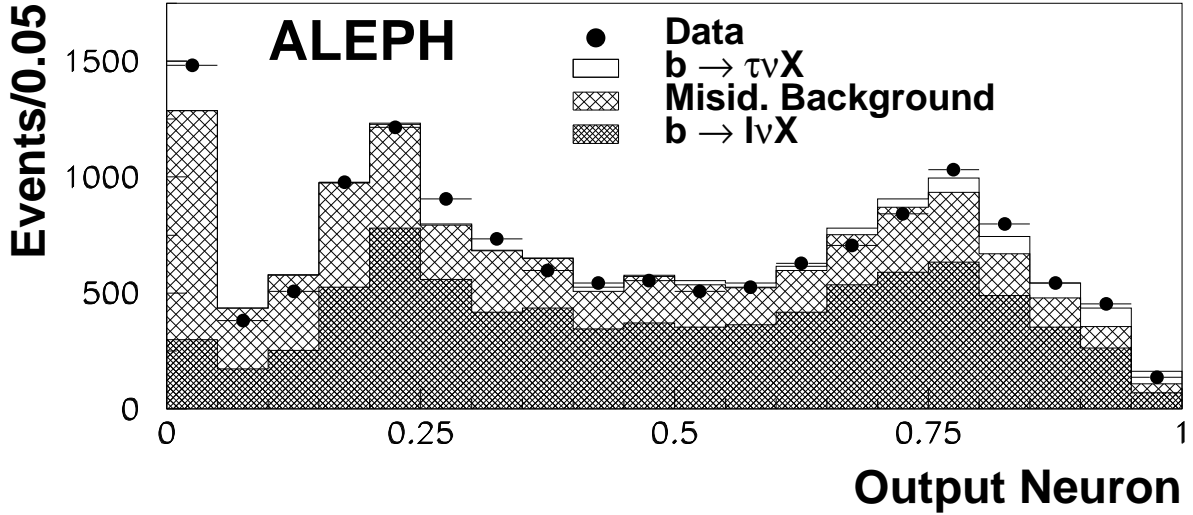}}
\end{picture}
\caption[ ]
{\protect\footnotesize Output neuron distributions for data and
simulation (histograms). The normalization of the 
simulation \bbtx ~distribution corresponds to the branching fraction 
value obtained from the fit. The samples labelled 
$\rm{b} \to \tau^- \bar{\nu}_{\tau} \rm{X}$ 
and $\rm{b} \to \ell \bar{\nu}_{\ell} \rm{X}$ 
in the figure contain only the  events where both leptons 
are correctly identified. 
Events with at least one lepton mis-identified are classified 
as ``Mis-identification background''.
\label{fig:out}}
\end{figure}

The agreement between data and \MC\ distributions is good for 
all the variables used in the analysis (Fig.~\ref{fig:comp}). 
The sample of simulated events is rescaled so that the number of 
selected hadronic events is equal to that in the data. The value of 
\bbtx\ is obtained by fitting the simulated output neuron distribution 
to the observed distribution, shown in Fig.~\ref{fig:out}. The 
discrepancy in the first bin has been traced, using the same sign 
di-lepton sample, to a subset of mis-identified muons not well reproduced 
by the simulation~\cite{tesis}. These background events are well 
separated from the signal events by the NN. The inclusion of
the first bin in the fit has a negligible effect on the fitted
branching fraction.

The dominant contributions to the systematic errors are the uncertainties 
in the values of the branching ratios involved in the analysis, 
the uncertainties in the modelling of semi-leptonic decays, 
the effect of the \emiss\ calibration previously discussed
and those inherent to the lepton identification. 
Some sources of systematic uncertainty (such as those related to the lepton 
identification) which affect both $\rm{b} \to \tau^- \bar{\nu}_\tau \rm{X}$ 
and $(\rm{b} \to \rm{c} \to \ell)$ are properly taken into account.
All the errors were obtained by varying the amount of the component 
of the event sample under study according to its uncertainty, changing the 
distribution of their input variables to the NN when required, and re-fitting 
the NN output neuron. The changes in the fitted fraction of 
$\rm{b} \to \tau^- \bar{\nu}_\tau \rm{X}$ events and in BR$(\rm{b} \to 
\rm{c} \to \ell)$ are evaluated with their respective NN, and the 
corresponding change in the BR($\rm{b} \to \tau^- \bar{\nu}_\tau \rm{X}$) 
is calculated.  

The result is 
$$\bbtx= \left[ \btxbm \tpm \btxbea\  \rm{(stat.)}\  \tpm \btxbeb\
  \rm{(syst.)} \right] \tpc.$$

Various consistency checks are performed. The lepton identification 
performance are tested by obtaining results separately for the three 
subsamples of di-electrons
($\rm{e}^\pm, \rm{e}^{\mp}$), di-muons
($\mu^\pm,\mu^{\mp}$) and ($\rm{e}^\pm,\mu^{\mp}$) pairs.
The measured branching fractions \bbtx\ are found to be 
$(3.8 \pm 1.3)\%$, $(3.6 \pm 1.4)\%$ and $(2.9 \pm 1.0)\%$,
respectively.  The simulation of the missing energy is tested in two 
ways. First, a NN fed with charged track information only gives
\bbtx\ =$(3.15\pm 0.89)\%$. Second, the branching fraction obtained 
when fitting the \emiss\ distribution instead of the NN output is 
(3.05 \tpm 0.90)\tpc.

\parskip 0.25cm plus 0.10cm
\section{Interpretation in two-Higgs-doublet models}
\label{sec:typeII}

As discussed in Section~\ref{sec:intro}, all five measurements
presented in this paper can in principle be used to constrain
type-II two-Higgs-doublet models. Practically, however, the constraints 
obtained in this framework are largely dominated by two of these measurements. 
\begin{itemize}
\item The Z penguin diagrams (Fig.~\ref{fig:graphes}f) are
related to photon penguin diagrams already  severely 
constrained by $\rm{b} \to \rm{s}\gamma$ searches, and cannot 
contribute significantly to \bts~\cite{bsnunu_theory}.
\item The \bbtd\ measurement is significantly correlated with,
and is statistically less powerful than that of \bbtx.
\item The \bbtx\ measurement with di-leptons is much less accurate 
than the corresponding measurement based on missing energy.
\end{itemize}

Consequently, only the \bbtx\ measurement based on missing energy and the 
search for the \bt\ final state, kept statistically independent in the 
analysis, were interpreted to constrain type-II two-Higgs-doublet models. 
In these models, both branching ratios depend on the parameter $r$,
$$r \equiv {\tan\beta \over \mHpm},$$
where \tanb\ is the ratio of the vacuum expectations values of the 
two Higgs doublets, and \mHpm\ is the mass of the resulting charged
Higgs boson. In the free-quark model, the \btx\ branching ratio is expected 
to be modified with respect to the standard model prediction according 
to~\cite{higgs_btaux0}
\begin{equation}
\label{eq:bbtx}
\bbtx = \bbtxSM \times \left[ 1 - 2 m_\tau^2 r^2 \Phi +  
{m_\tau^2 m_{\rm{b}}^2 \over 4} r^4 \right],
\end{equation}
where $\Phi$ is a phase-space factor depending on $m_\tau$, $m_{\rm{c}}$ and
$m_{\rm{b}}$ amounting to about 0.6. An enhancement can therefore be observed
for values of $r$ in excess of $\sim 0.43\,(\Gcs)^{-1}$, while the destructive 
interference yields a reduction of the branching ratio below that value. 
Similarly, the \bt\ fraction is modified with respect to the standard model by
\begin{equation}
\label{eq:bbt}
\bbt = \bbtSM \times \left[ 1 - r^2 m_{{\mathrm{B}}^-}^2   \right]^2.
\end{equation}
which shows an even stronger dependence on $r$, and represents an
enhancement for all values of $r$ larger than $\sim 0.27\,(\Gcs)^{-1}$.

To extract a limit on $r$, the neutrino energy spectrum and the $\tau$ 
polarization were computed as a function of $r$~\cite{higgs_btaux1}, 
and the simulated events, generated with the standard model values, were 
re-weighted accordingly. The \bbtx\ measurement and the \bbt\ limit were 
derived as presented in Sections~\ref{sec:meast-bbtx} and~\ref{sec:limits}, 
for any $r$ value. 

From the dependence between \bbtx\ and $r$ of Eq.~\ref{eq:bbtx}, 
and from the value of \bbtxSM, a 90\% C.L. upper limit on $r$ was 
extracted~\cite{higgs_btaux1,higgs_btaux2,limitprivate1}:
\begin{equation}
\label{eq:btxlim}
\tanbomh < \tbomhm \left(\Gcs\right)^{-1} \cl.
\end{equation}
The $\tau$ polarization actually plays here an important r\^ole since, for 
$r=0.49\,(\Gcs)^{-1}$, it amounts to $-0.28$ while the standard model 
prediction is $-0.735$ in \btx\ decays. This effect had been neglected 
in previous ALEPH analyses~\cite{aleph_btaux2}.

The dependence between \bbt\ and $r$ of Eq.~\ref{eq:bbt} and
the standard model value of the branching ratio, 
$$\bbtSM= 7.4 \times 10^{-5} \left(f_{\mathrm{B}}/160\,\rm{MeV}\right)^2 
\left(\vert \mathrm{V}_{\mathrm{ub}} \vert /0.004\right)^2,$$ 
were interpreted in terms of a 90\% C.L. upper limit on $r$:
\begin{equation}
\label{eq:btlim}
\tanbomh < 0.40 \left(\Gcs\right)^{-1} \cl.
\end{equation}
In the limit extraction, the systematic uncertainties on 
$f_{\mathrm{B}}$~\cite{fB} and on $\mathrm{V}_{\mathrm{ub}}$~\cite{Vub} were 
taken into account following the method of Ref.~\cite{cousins}. The 
combination of the two results in Eqs.~\ref{eq:btxlim} and~\ref{eq:btlim}
cannot improve on the latter, since the corresponding $r$ value does not 
enhance the \btx\ branching ratio.

\section{Conclusion} 
\label{sec:conclusion}
With approximately four million hadronic Z decays collected by the 
ALEPH detector at LEP, branching ratios involving a $\rm{b} \to \tau$
transition have been  measured to be
\begin{eqnarray*}
\bbtx &=& (\btxam \pm \btxaea \pm \btxaeb)\tpc, \\
\bbtd &=& (\btdm \pm \btdea \pm \btdeb)\tpc, 
\end{eqnarray*}
in agreement with the standard model predictions and consistent with 
similar measu\-rements performed by DELPHI~\cite{delphi_btaux1} and
L3~\cite{l3_btaux1,l3_btaux2}.
A search for the exclusive decay \bt\ has allowed a 90\% C.L. upper 
limit to be set on the corresponding branching ratio,
$$\bbt < \btm \cl.$$
Similar limits have been obtained by CLEO~\cite{btau_cleo}, L3~\cite{l3_btaux3}
and DELPHI~\cite{delphi_btaux1}. In the framework of type-II 
two-Higgs-doublet models, these results translate to a constraint 
on the model parameter $r = \tanbomh$,
$$\tanbomh < 0.40 \left(\Gcs\right)^{-1} \cl.$$
Finally, a limit has been set on the process \bts\ to
$$\bbts < \btsma \cl.$$
A related limit on ${\mathrm{BR(B}}^- \to {\mathrm{K}}^-\nu\bar\nu)$ has 
been obtained by CLEO~\cite{btau_cleo}.

\section*{Acknowledgments}

We would like to thank Yuval Grossman, Zoltan Ligeti and Enrico Nardi 
for fruitful discussions on the \bts\ process and the computation of 
the limit on \tanbomh\ in the \btx\ channel. We are indebted to our 
colleagues in the accelerator divisions for the good performance of LEP.
We thank also the engineers and technicians of all our institutions
for their contribution to the excellent performance of ALEPH.
Those of us from non-member countries thank CERN for its hospitality.


\end{document}